\documentclass[twocolumn]{aastex631}

\received{}
\revised{}
\accepted{}
\submitjournal{}

\shorttitle{Dense molecular gas in merger remnants}
\shortauthors{Ueda et al.}
\begin{document}

\title{COLD MOLECULAR GAS IN MERGER REMNANTS. II. 
The properties of dense molecular gas}

\correspondingauthor{Junko Ueda}
\email{junko.ueda@nao.ac.jp}

\author[0000-0003-3652-495X]{Junko Ueda}
\affiliation{
National Astronomical Observatory of Japan, National Institutes of Natural Sciences, 
2-21-1 Osawa, Mitaka, Tokyo, 181-8588, Japan}

\author[0000-0002-2364-0823]{Daisuke Iono}
\affiliation{
National Astronomical Observatory of Japan, National Institutes of Natural Sciences, 
2-21-1 Osawa, Mitaka, Tokyo, 181-8588, Japan}
\affiliation{
Department of Astronomical Science, The Graduate University for Advanced Studies, SOKENDAI, 
2-21-1 Osawa, Mitaka, Tokyo 181-8588, Japan}

\author[0000-0001-7095-7543]{Min S. Yun}
\affiliation{
Department of Astronomy, University of Massachusetts, Amherst, MA 01003, USA}

\author[0000-0003-2475-7983]{Tomonari Michiyama}
\affiliation{
National Astronomical Observatory of Japan, National Institutes of Natural Sciences, 
2-21-1 Osawa, Mitaka, Tokyo, 181-8588, Japan}
\affiliation{
Department of Earth and Space Science, Graduate School of Science, Osaka University, 
1-1 Machikaneyama, Toyonaka, Osaka 560-0043, Japan}

\author[0000-0002-9668-3592]{Yoshimasa Watanabe}
\affiliation{
Materials Science and Engineering, College of Engineering, Shibaura Institute of Technology,
3-7-5 Toyosu, Koto-ku, Tokyo 135-8548, Japan}

\author{Ronald L. Snell}
\affiliation{
Department of Astronomy, University of Massachusetts, Amherst, MA 01003, USA}

\author[0000-0003-1327-0838]{Daniel Rosa-Gonz\'{a}lez}
\affiliation{
Instituto Nacional de Astrof\'{i}sica, \'{O}ptica y Electr\'{o}nica, 
Luis Enrique Erro 1, Tonantzintla, Puebla, C.P. 72840, M\'{e}xico}

\author[0000-0002-2501-9328]{Toshiki Saito}
\affiliation{
National Astronomical Observatory of Japan, National Institutes of Natural Sciences, 
2-21-1 Osawa, Mitaka, Tokyo, 181-8588, Japan}
\affiliation{
College of Engineering, Nihon University, 
1 Nakagawara, Tokusada, Tamuramachi, Koriyama, Fukushima 963-8642, Japan}

\author[0000-0002-2852-9737]{Olga Vega}
\affiliation{
Instituto Nacional de Astrof\'{i}sica, \'{O}ptica y Electr\'{o}nica, 
Luis Enrique Erro 1, Tonantzintla, Puebla, C.P. 72840, M\'{e}xico}

\author[0000-0002-4999-9965]{Takuji Yamashita}
\affiliation{
National Astronomical Observatory of Japan, National Institutes of Natural Sciences, 
2-21-1 Osawa, Mitaka, Tokyo, 181-8588, Japan}
\affiliation{
Research Center for Space and Cosmic Evolution, Ehime University, 
2-5 Bunkyo-cho, Matsuyama, Ehime 790-8577, Japan}

\begin{abstract}
We present the 3~mm wavelength spectra of 28 local galaxy merger remnants 
obtained with the Large Millimeter Telescope. 
Fifteen molecular lines from 13 different molecular species and isotopologues were identified, 
and 21 out of 28 sources were detected in one or more molecular lines. 
On average, the line ratios of the dense gas tracers, such as HCN~(1--0) and HCO$^{+}$(1--0), 
to $^{13}$CO~(1--0) are 3--4 times higher in ultra/luminous infrared galaxies (U/LIRGs) 
than in non-LIRGs in our sample. 
These high line ratios could be explained by the deficiency of $^{13}$CO 
and high dense gas fractions suggested by high HCN~(1--0)/$^{12}$CO~(1--0) ratios. 
We calculate the IR-to-HCN~(1--0) luminosity ratio 
as a proxy of the dense gas star formation efficiency. 
There is no correlation between the IR/HCN ratio and the IR luminosity, 
while the IR/HCN ratio varies from source to source 
(1.1--6.5) $\times 10^{3}$ $L_{\sun}$/(K\,km\,s$^{-1}$\,pc$^{2}$). 
Compared with the control sample, we find that the average IR/HCN ratio of the merger remnants 
is higher by a factor of 2--3 than those of the early/mid-stage mergers and non-merging LIRGs, 
and it is comparable to that of the late-stage mergers. 
The IR-to-$^{12}$CO~(1--0) ratios show a similar trend to the IR/HCN ratios. 
These results suggest that star formation efficiency is enhanced by the merging process 
and maintained at high levels even after the final coalescence. 
The dynamical interactions and mergers could change the star formation mode 
and continue to impact the star formation properties of the gas in the post-merger phase.
\end{abstract}


\section{Introduction}

Dynamical interactions and mergers between gas-rich galaxies can trigger 
dust-obscured starbursts, resulting in galaxies bright in infrared (IR) luminosity 
\citep[U/LIRGs ($10^{11}\,L_{\sun} \leq L_{\rm IR} < 10^{13}\,L_{\sun}$);][]{1996ARA&A..34..749S}. 
Simulations predict that the merger-driven starburst peaks at the final coalescence 
and that an active galactic nucleus (AGN) is triggered in some cases 
\citep[e.g.,][]{1996ApJ...464..641M, 2008ApJS..175..356H}. 
An increasing number of observational studies utilizing surveys have investigated 
how the star formation activity varies across different stages of mergers 
\citep[e.g.,][]{2008AJ....135.1877E, 2008MNRAS.385.1903L, 2016PASJ...68...96M, 2018MNRAS.476.2591V}. 
By studying a sample of 58 galaxy pairs, 
\citet{2018ApJ...868..132P} find that the star formation rate (SFR) increases 
with decreasing separation of galaxies in pairs. 
In addition, by using a sample of $>$10000 galaxies identified in the Sloan Digital Sky Survey, 
\citet{2013MNRAS.435.3627E} reveal that the fraction of starburst galaxies 
peaks in the post-merger phase, where the two nuclei of the interacting galaxies have merged.

Active star formation occurs in dense regions within molecular clouds. 
Simulations predict that the amount of dense gas significantly increases during the merger, 
leading to a high fraction of dense gas \citep{2009ApJ...707.1217J, 2019MNRAS.485.1320M}. 
This is considered to be one of the main factors responsible for merger-driven starbursts. 
The simulated probability distribution function of gas density 
in a starbursting major merger strongly evolves toward high densities 
during the merging process due to the rapidly increased turbulence 
and numerous local shocks compressing the gas \citep{2011EAS....51..107B}. 
This excess should have signatures in observable molecular line ratios 
(e.g., HCN~(1--0)/$^{12}$CO~(1--0)). 
While $^{12}$CO~(1--0) (hereafter CO~(1--0)) is the best tracer of diffuse molecular gas 
because of its low critical density ($n_{\rm crit} \sim 10^{2.5}$~cm$^{-3}$) 
and relatively high abundance, other less abundant molecular species, 
such as HCN, HCO$^{+}$ and HNC, can be the dense gas tracers. 
Since these molecules have larger dipole moments than CO, 
they require higher densities for collisional excitation. 
Thus, the emission from these molecules traces dense molecular gas 
most directly linked to star formation. 
This is supported by the tight linear correlation between the HCN~(1--0) luminosity 
and the IR luminosity \citep[e.g.,][]{2004ApJ...606..271G, 2005ApJ...635L.173W, 2017A&A...604A..74S}.

Another factor for merger-driven starbursts is 
an increase in the efficiency of converting gas into stars. 
This efficiency (star formation efficiency = SFE) of galaxies 
has been discussed using the ratio between SFR and the molecular gas mass ($M_{\rm H_{2}}$) 
or the dense molecular gas mass ($M_{\rm dense}$). 
The tight correlation between the HCN and IR luminosities \citep{2004ApJ...606..271G} 
indicates that the dense gas star formation efficiency (SFE$_{\rm dense}$ = SFR/$M_{\rm dense}$) 
would be approximately constant for all galaxies. 
However, \citet{2012A&A...539A...8G} find that the SFE$_{\rm dense}$ of U/LIRGs is 
a few times higher than the SFE$_{\rm dense}$ of normal galaxies. 
As their sample of U/LIRGs includes merging galaxies, 
the SFE$_{\rm dense}$ of mergers is likely enhanced. 
On the other hand, \citet{2018MNRAS.476.2591V} find that the depletion time ($t_{\rm dep}$ = 1/SFE) 
of galaxy pairs is consistent with $t_{\rm dep}$ of non-mergers with similarly elevated SFRs, 
suggesting that galaxy interactions and mergers do not enhance the SFE. 
\citet{2018ApJ...868..132P} obtained a similar result that the SFE does not increase 
with decreasing separation of pair galaxies, 
but they also find a signature for the SFE enhancement 
in close pairs (separation $<$ 20~kpc) and equal-mass systems. 
Furthermore, the SFE and SFE$_{\rm dense}$ seem to vary within a galaxy. 
\citet{2019AJ....157..131B} find that the total IR luminosity to the HCN luminosity ratio 
varies by up to a factor of $\sim$10 across different regions of the mid-stage merger NGC~4038/9.
A conclusion has not been reached about 
how SFE and SFE$_{\rm dense}$ change throughout the merging process.

We thus carried out dense gas observations towards 28 merger remnants 
using the Large Millimeter Telescope \citep[LMT;][]{2010SPIE.7733E..12H} 
to characterize the properties of dense molecular gas using the line ratios 
and investigate the star formation properties of the gas in the final stages of mergers. 
Merger remnants are completely merged galaxies 
that still have tidal tails, shells, and loops, 
which indicate past dynamical interactions. 
In this study, we measure the HCN~(1--0)/CO~(1--0), IR-to-CO~(1--0), 
and IR-to-HCN~(1--0) luminosity ratios as proxies for the dense gas fraction, 
SFR per unit molecular gas mass, 
and SFR per unit dense gas mass (SFE$_{\rm dense}$), respectively. 
In addition, we investigate the local environments of the gas 
using the line ratio between the dense gas tracers (HCN/HCO$^{+}$ and HNC/HCN). 
The HCN/HCO$^{+}$ ratio is suggested as a diagnostic 
for cosmic rays \citep{2011A&A...525A.119M}, 
X-ray dominated region (XDR) \citep{2005A&A...436..397M}, 
photon dominated region (PDR) \citep{2012A&A...542A..65K}, 
and mechanical heating \citep{2008A&A...488L...5L}. 
High HCN/HCO$^{+}$ line ratios can be a tracer of AGN-dominated systems 
\citep{2001ASPC..249..672K, 2007AJ....134.2366I} 
because the high line ratios could be explained by an enhanced HCN abundance 
in XDRs surrounding an AGN. 
The isomer abundance ratio between HNC and HCN might serve 
as an indicator of PDR and XDR \citep{2008A&A...477..747B}, 
although this line ratio could be increased by IR radiation pumping \citep{2007A&A...464..193A}.

This paper is organized as follows. 
We present our sample, observational details, and data reduction process in Section~2. 
Then we show new results in Section~3. 
In Section~4, we investigate the average properties of molecular gas 
in the merger remnants by stacking the spectra of our sample. 
We discuss the line ratios between dense gas tracers, dense gas fraction, 
and star formation efficiency while comparing the control samples in Section~5 and 6. 
We summarize this paper in Section~7. 
All values of our sample sources are calculated based on a $\Lambda$CDM model 
with $H_{0}$ = 73 km s$^{-1}$ Mpc$^{-1}$, $\Omega_{\rm M}$ = 0.3, 
and $\Omega_{\rm \Lambda}$ = 0.7, which are the same values 
used in the first paper of this series \citep{2014ApJS..214....1U}.

\section{Observations and ancillary data}

\subsection{Sample Sources}
The sample used in this study is a subset of 
the CO imaging data of merger remnants \citep{2014ApJS..214....1U}. 
Our sample is originally drawn from 51 optically-selected merger remnants 
in the local ($<$200~Mpc) universe \citep{2004AJ....128.2098R}. 
The merger remnants were selected solely based on optical morphology 
that suggests an advanced merger stage, 
regardless of the strength of the starburst or AGN activities. 
Near-IR imaging shows single-nuclei in all of the sample sources \citep{2004AJ....128.2098R}, 
however radio continuum maps of NGC~3256 \citep{1995ApJ...446..594N} reveal 
a double nucleus, which suggests an ongoing merger. 
In this study, we regard NGC~3256 as a merger remnant. 
The basic properties of our sample are summarized in Table~\ref{tab:t1}. 
The exact fraction of these sources that result from a major merger 
as opposed to a minor merger is unknown 
because it is difficult to reverse the chronology 
and disentangle the exact mass and morphology of the progenitors. 
We estimated the far-IR (FIR) and IR luminosities of our sample using the \textit{IRAS} catalogs 
and Equation in Table~1 of \citet{1996ARA&A..34..749S}. 
The IR luminosities including the upper limits range 
from $1.4 \times 10^{9}~L_{\sun}$ to $3.3 \times 10^{12}~L_{\sun}$. 
Two galaxies are classified as ULIRGs 
($10^{12}\,L_{\sun} \leq L_{\rm IR} < 10^{13}\,L_{\sun}$), 
and 11 galaxies are classified as LIRGs 
($10^{11}\,L_{\sun} \leq L_{\rm IR} < 10^{11}\,L{\sun}$)
(see Table~\ref{tab:t1}).

\subsection{Observations with LMT}
Multi-line observations towards 28 merger remnants were carried out using the LMT 
between October 2014 and May 2015 in its early science phase. 
In this early phase, the LMT operated with a 32-m active surface. 
The Redshift Search Receiver \citep[RSR;][]{2007ASPC..375...71E} consists of 
two dual-polarization front-end receivers 
that are chopped between the ON and OFF source positions 
separated by 76\arcsec~in Azimuth, always integrating on-source. 
The back-end spectrometer covered the frequency range 73~GHz to 111~GHz simultaneously. 
The spectral resolution is 31.25~MHz, which corresponds to 100~km~s$^{-1}$ at 93~GHz. 
During the Early Science phase operation, 
only the inner 32~m diameter section of the telescope surface was illuminated. 
The primary beam size is 20\arcsec at 110~GHz and 28\arcsec at 75~GHz.

Data reduction was carried out using DREAMPY 
(Data REduction and Analysis Methods in PYthon), 
which is the pipeline software for the RSR data reduction written by G. Narayanan. 
After flagging scans affected by hardware or software problems, 
a linear baseline is removed from each spectrum. 
The rms noise level of the spectra at frequencies below 80~GHz 
are much higher than that for higher frequencies. 
There are no strong molecular lines in this frequency range. 
We thus excluded this frequency range and estimated the rms noise level of each spectrum 
using line-fee channels between $\nu_{\rm obs}$ = 80~GHz and 111~GHz. 
The rms noise levels in the frequency resolution of 31.25~GHz range 
between 0.14~mK and 0.40~mK (Avg. = 0.29 $\pm$ 0.01~mK) 
in the antenna temperature ($T_{\rm A}^{*}$) unit.

We used the IDL's GAUSSFIT function to identify molecular lines. 
We consider molecular lines whose peak intensity can be identified 
with a signal-to-noise ratio (S/N) of $>$ 3 as ``detected lines''. 
Molecular lines detected with 2 $<$ S/N $<$ 3 are considered 
to be ``tentatively-detected line'' 
if their velocities and line widths agree with 
those of other detected lines in the same sources. 
The Gaussian line fitting failed for C$_{2}$H~(1--0) 
as this feature is a blend of two transitions, each with hyperfine structure. 
Therefore for C$_{2}$H~(1--0) we consider the line to be detected 
if at least two spectral channels at the correct frequency are above 3$\sigma$.

\subsection{Ancillary data: ALMA CO maps}
We used the CO~(1--0) data of six sources 
(Arp~187, AM~0956-282, NGC~3597, AM~1300-233, AM~2055-425, and NGC~7252) 
obtained with the Atacama Large Millimeter/submillimeter Array (ALMA) 12~m array and 
the Atacama Compact Array (ACA: 7~m array + Total Power (TP) array) 
as part of 2011.0.00099.S, 2016.2.00006.S, and 2017.1.01003.S. 
In these projects, one spectral window was set to cover the redshifted CO~(1--0) line. 
The bandwidth of the spectral window is 1.875~GHz 
and the frequency resolution is 488 kHz. 
In addition, we used the CO~(1--0) data of NGC~3256 obtained 
with the ALMA 12~m (TM2) array and the ACA 7~m array 
as part of 2016.2.00042.S, 2016.2.00094.S, and 2018.1.00223.S. 
The data were obtained using a single pointing.
The full width at half maximum (FWHM) of the primary beam is $\sim$51\arcsec 
at the observing frequency for a 12~m antenna. 
We adopt the typical systematic errors on the absolute flux calibration of 5\% 
for the Band 3 data \citep[e.g.,][]{THCy0, PGCy4}.

We first restored the calibrated measurement set 
using the observatory-provided reduction script and the apropriate versions of 
Common Astronomy Software Applications (CASA) package. 
Flux re-scaling was applied to the ACA 7~m array data of Arp~187, NGC~3597, and AM~1300-233 
because the more appropriate catalogue values of the amplitude calibrators are available 
in the ALMA Calibrator Source Catalogue. 
After the continuum subtraction, we combined the 12~m array data and the 7~m array data 
and created the 12m+7m images by adopting Briggs weighting of the visibilities (robust= 0.5). 
For 6/7 sources, we also made the TP images 
and combined them with the 12m+7m images using the CASA feather task. 
The data of NGC~3256 were not corrected with the zero-spacing information 
because the TP array data are not available. 
The maximum recoverable angular scale calculated from the minimum baseline 
of the 7~m array ($\sim$9~m) is $\sim$35\farcs8 at the observing frequency. 
We compared the flux density measured in the 12m+7m image of NGC~3256 
with the literature single-dish measurement \citep{1992A&A...264...49C}. 
The recovered flux is 90\% in the central 44\arcsec region. 
The image rms per channel and the synthesized beam size are summarized in Table~\ref{tab:t2}.

\section{Results}
Twenty-one of 28 merger remnants were detected with S/N $>$~3 in one or more molecular lines.
The detection rate does not depend on the distances to galaxies.
The spectra of the merger remnants are presented in Figure~\ref{fig:f1}. 
The molecular lines in NGC~3256 are roughly five times brighter 
than the galaxy with second brightest lines.
Of the 21 galaxies detected in one or more molecular lines, 
all but UGC~10675 was detected in the $^{13}$CO~(1--0) line.
Seventeen sources were detected in either the HCN~(1--0) or HCO$^{+}$(1--0) lines 
and 15 were detected in both lines. 
While UGC~10675 was detected with S/N $\sim$ 5 in the HCN line, 
it was not detected in the HCO$^{+}$ line.
On the other hand, NGC~6052 was detected with S/N $>$ 6 in the HCO$^{+}$ line, 
but it was not detected in the HCN line.
Six out of 15 sources have the HCN luminosity higher than the HCO$^{+}$ luminosity.
The HC$_{3}$N lines were detected in UGC~5101 and UGC~8058, 
both of which host X-ray detected AGNs \citep[e.g.,][]{2011A&A...529A.106I}.
Three molecular species (c-C$_{3}$H$_{2}$, CH$_{3}$OH, and CH$_{3}$CCH) 
were only detected in NGC~3256 and N$_{2}$H$^{+}$ was only detected in UGC~8058.
The peak intensities of these species are weaker compared to the major species 
such as $^{13}$CO, HCN, and HCO$^{+}$.
In addition, the C$_{2}$H~(1--0), HNC~(1--0), C$^{18}$O~(1--0), and CS~(2--1) lines 
were identified in several sources.
There are two transitions of C$_{2}$H ($N$ = 1--0 $J$ = 3/2--1/2 and $N$ = 1--0 $J$ = 1/2--1/2) 
in the frequency range, but these transitions cannot be resolved 
due to a limited frequency resolution.  
For relatively distant sources ($D_{\rm L} \geq$ 90~Mpc), 
redshifted CN ($N$ = 1--0, $J$ = 1/2--1/2), CN ($N$ = 1--0, $J$ = 3/2--1/2), 
and CO~(1--0) lines fell into the bandwidth.
The CN lines were detected in six sources, and the CO line was detected in two sources.
The hyperfine lines of CN cannot be resolved due to the limited frequency resolution.
In summary, 15 molecular lines from 13 different molecular species and isotopologues 
were identified within the frequency range.

The properties of molecular lines identified, including the integrated line intensity, 
are summarized in Table~\ref{tab:t3}.
We estimated the integrated line intensity 
by summing consecutive channels ($N_{\rm ch}$) whose values are above 1.5$\sigma$ 
around the central channel identified by the Gaussian fitting.
We did not use the results of the Gaussian fitting 
to estimate the integrated line intensities 
because some of the molecular lines show double horn line profiles 
even if they can be fitted with Gaussian profiles. 
In the case of non-detection in six major species 
(HCN, HCO$^{+}$, HNC, CS, C$^{18}$O, and $^{13}$CO), 
we estimated the 3$\sigma$ upper limit of the integrated line intensity 
by using the following equation, 
\begin{equation}
I_{\rm 3\sigma}=3\times{\rm RMS}\times\Delta\,V_{\rm ch}\sqrt{N_{\rm ch}},
\end{equation}
where RMS is the 1$\sigma$ noise level of the spectrum, 
$\Delta\,V_{\rm ch}$ is the velocity width of a channel at the observing frequency, 
and $N_{\rm ch}$ is the number of channels integrated.
We used $N_{\rm ch}$ of $^{13}$CO to estimate the upper limit of C$^{18}$O 
and the average $N_{\rm ch}$ of detected lines among HCN, HCO$^{+}$, HNC, and CS 
to estimate the upper limit of the other lines.
When none of these four lines were detected, 
we used the number of channels to cover the line width ($\Delta\,V \times N_{\rm ch}$) 
of $^{13}$CO to estimate the upper limits of the four lines.
For UGC~10675, we estimated the upper limits of HCO$^{+}$, HNC, CS, C$^{18}$O, and C$^{13}$O 
using the number of channels to cover the line width of HCN.
For seven galaxies which were not detected in any molecular lines, 
the 3$\sigma$ upper limit of the integrated line intensity was estimated, 
assuming that the line width is similar to that of the $^{12}$CO~(1--0) emission 
\citep{2014ApJS..214....1U}.
We use a Kelvin--to--Jansky gain factor of 7 Jy K$^{-1}$ (in $T_{\rm A}^{*}$ unit) 
to convert the units of line intensity to Jansky, 
assuming that the source is smaller than the beam.
Then, we calculated the luminosities of HCN, HCO$^{+}$, and HNC 
in units of K~km~s$^{-1}$ pc$^{2}$ using the equation (3) in \citet{2005ARA&A..43..677S}:
\begin{equation}
L'_{\rm line}=3.25\times10^{7}S_{\rm line}\Delta\,v\nu_{\rm obs}^{-2}D_{\rm L}^{2}(1 + z)^{-3},
\end{equation}
where $L'_{\rm line}$ is the luminosity in the unit of K km s$^{-1}$ pc$^{2}$, 
$S_{\rm line}\Delta\,v$ is the integrated flux density in Jy km s$^{-1}$,
$\nu_{\rm obs}$ is the observing frequency in GHz, 
$D_{\rm L}$ is the luminosity distance in Mpc, 
and $z$ is the redshift.
The derived luminosities are summarized in Table~\ref{tab:t4}.
We note that the data of UGC~5101 has been published in \citet{2020MNRAS.499.2042C}, 
but they combined additional data obtained in 2017.
Ten molecular lines were identified in the spectrum of UGC~5101.
The derived integrated intensities of 8/10 molecular lines are consistent with 
the values reported by the previous study within the errors.
The integrated intensities of C$_{2}$H~(1--0) and CN~(1--0; 3/2--1/2) 
are 73\% and 89\% of the previous measurements \citep{2020MNRAS.499.2042C}.
These lines are blended with the hyperfine lines and broadened, 
but our analysis may miss the faint hyperfine lines whose peaks are below 1.5$\sigma$.

\section{Line ratios of dense gas tracers to $^{13}$CO}
\subsection{Dependence on the IR luminosity}
We calculate the intensity ratios of the dense gas tracers 
(HCN~(1--0), HCO$^{+}$(1--0), and HNC~(1--0)) to $^{13}$CO~(1--0) 
of 20 merger remnants detected in $^{13}$CO~(1--0). 
The intensity ratios are shown as a function of the IR luminosity in Figure~\ref{fig:f2}. 
Eleven out of 20 sources are U/LIRGs ($L_{\rm IR} \ge 10^{11}\,L_{\sun}$), 
and the remaining nine sources are non-LIRGs ($L_{\rm IR} < 10^{11}\,L_{\sun}$). 
While the HCN/$^{13}$CO and HCO$^{+}$/$^{13}$CO intensity ratios are higher than 
the unity for most of the U/LIRGs, they are lower than the unity for the non-LIRGs. 
On average, the HCN/$^{13}$CO and HCO$^{+}$/$^{13}$CO intensity ratios of the U/LIRGs 
are 4.2 and 2.8 times higher than the non-LIRGs, respectively, 
suggesting different interstellar medium (ISM) environments.
The HCN/$^{13}$CO and HCO$^{+}$/$^{13}$CO intensity ratios of one ULIRG (UGC~8058) 
are higher than 10, which is one order of magnitude higher than those of the other U/LIRGs.
Even when we exclude UGC~8058, the average HCN/$^{13}$CO and HCO$^{+}$/$^{13}$CO ratios 
of the remaining ten U/LIRGs are 2.1 and 1.7 times higher than those of the non-LIRGs, respectively.
Since all the HNC/$^{13}$CO intensity ratios of the non-LIRGs are upper limits, 
it is difficult to investigate its dependency on the IR luminosity.

One possibility for the elevated HCN/$^{13}$CO and HCO$^{+}$/$^{13}$CO intensity ratios 
is a high dense gas fraction. 
Since the critical densities of HCN~(1--0) and HCO$^{+}$(1--0) are 
2--3 orders of magnitude higher than that of $^{13}$CO~(1--0) \citep{2005pcim.book.....T}, 
the HCN~(1--0) and HCO$^{+}$ emission can trace the dense molecular gas. 
These elevated line ratios imply high dense gas fractions in the U/LIRGs, 
which are also suggested by high HCN~(1--0)/$^{12}$CO~(1--0) ratios that we derive in \S 6.1. 
Another possibility is IR radiative pumping. 
HCN can be vibrationally excited to $v_{2}$ = 1 
by absorbing mid-IR (14~$\mu$m) photons \citep{1981ApJ...245..891C}. 
Then, HCN can decay back to the vibrational-ground level ($v$ = 0).
The subsequent cascade from the $J$=2 state can enhance 
the intensity of the HCN~(1--0) emission 
compared to the case when the excitation is caused only by collisions 
\citep{2011ApJ...743...94R, 2010ApJ...725L.228S}.
However, several studies have concluded that the IR pumping is not 
the main mechanism to explain high HCN luminosities 
\citep[e.g.,][]{2004ApJS..152...63G, 2016AJ....152..218I}.
HCO$^{+}$ has a similar vibrational bending state at 12~$\mu$m, 
but the excitation of HCO$^{+}$ through IR radiative pumping 
is less than that of HCN \citep{2016AJ....152..218I}.
In this study, both HCO$^{+}$/$^{13}$CO and HCN/$^{13}$CO ratios are enhanced in the U/LIRGs. 
The elevated line ratios can be partially but not fully explained by only IR pumping. 
A deficiency of $^{13}$CO is another possibility for the elevated line ratios in the U/LIRGs.
It is known that the $^{13}$CO emission is unusually weak relative to $^{12}$CO in U/LIRGs.
The $^{12}$CO/$^{13}$CO line ratios exceed 20 and reach 50 for some sources 
\citep{1991A&A...249..323A, 2017ApJ...840L..11S}, 
which are much higher than the $^{12}$CO/$^{13}$CO line ratio of disk galaxies 
\citep[$\sim$10;][]{2001ApJS..135..183P}. 
Several scenarios can explain high $^{12}$CO/$^{13}$CO ratios, 
including optical depth effects and abundance variations
\citep[e.g.,][]{1992A&A...264...55C, 1993A&A...274..730H}.

\subsection{Dependence on the CO extent}
We check the relationship between the spatial extent of molecular gas 
obtained from previous high-resolution CO imaging 
and the intensity ratios calculated in \S 4.1.
\citet{2014ApJS..214....1U} find that about half of the merger remnants 
have extended disks of molecular gas traced by the CO 
and suggest that those extended gas disks can rebuild stellar disks 
and form disk-dominated late-type galaxies (LTGs). 
On the other hand, the merger remnants with compact gas disks 
are candidates that could become early-type galaxies (ETGs).

In Figure~\ref{fig:f3}, we plot the intensity ratios of 17 sources 
as a function of the size of the molecular gas disk relative to the stellar component 
($R_{\rm ratio}=R_{80}/R_{\rm eff}$ in Table~\ref{fig:f1}).
$R_{80}$ is the radius from the galactic center, which contains 80\% of the CO flux, 
and $R_{\rm  eff}$ is the radius of the isophote 
containing half of the total $K$-band luminosity.  
Seven out of the 17 merger remnants with CO maps
have molecular gas disks with $R_{\rm ratio} > 1$, 
and ten sources have molecular gas disks with $R_{\rm ratio} < 1$. 
In general, the HCN/$^{13}$CO, HCO$^{+}$/$^{13}$CO, and HNC/$^{13}$CO ratios 
do not depend on $R_{\rm ratio}$. 
One galaxy (UGC~8058) stands out as having both exceptionally large $R_{\rm ratio}$ 
and molecular line ratios and as the most IR luminous galaxy in our sample. 
When we exclude UGC 8058, the average HCN/$^{13}$CO ratio of sources 
with $R_{\rm ratio}>1$ (0.92 $\pm$ 0.29) 
is comparable to that of sources with $R_{\rm ratio}<1$ (0.92 $\pm$ 0.15). 
Thus, the HCN/$^{13}$CO intensity ratios are not likely to be affected 
by the different evolutionary pathways predicted by the CO spatial extent. 
This is suggested from the comparison of the HCN/$^{13}$CO between ETGs and LTGs. 
The HCN/$^{13}$CO intensity ratios of ETGs range from 0.18 to 0.58 
\citep{2010MNRAS.407.2261K, 2012MNRAS.421.1298C}.
They are consistent with the HCN/$^{13}$CO intensity ratios of 
spiral galaxies and dwarf galaxies with normal SFRs
\citep[e.g.,][]{2010MNRAS.407.2261K, 2010PASJ...62..409M, 2014ApJ...788....4W}. 
No significant difference between the different morphological types of galaxies 
(i.e., ETG and LTG) has been found.

\section{Line Ratios between dense gas tracers}
In the following sections, we use literature data of local U/LIRGs 
in the subsample of the Great Observatories All-sky LIRG Survey (GOALS) 
\citep{2009PASP..121..559A, 2015ApJ...814...39P, 2019A&A...628A..71H} 
in order to compare the dense gas properties among different merger stages.
The local U/LIRGs are classified into early-stage mergers, 
mid-stage mergers, late-stage mergers, and non-merging LIRGs 
by visual inspection of the IRAC 3.6~$\mu$m images \citep{2013ApJS..206....1S}.
The late-stage merges may include post-mergers (i.e., merger remnants).
They are also classified into three groups 
(AGN-dominated systems, composite, and SB-dominated systems) 
based on the equivalent width of the 6.2~$\mu$m polycyclic aromatic hydrocarbon 
\citep{2015ApJ...814...39P}.
While there are no AGN-dominated systems in 13 early/mid-stage mergers and 19 non-merging LIRGs, 
four out of 12 late-stage mergers are AGN-dominated systems.
We also use literature data of LTGs in the sample of \citet{2004ApJ...606..271G} 
as the control sample.

\subsection{HCN~(1--0)/HCO$^{+}$(1--0) ratio}
It is proposed that a high HCN/HCO+ line ratio can be a tracer of AGN-dominated systems 
\citep{2001ASPC..249..672K, 2007AJ....134.2366I}, 
although the line ratio changes depending on the observed angular scale. 
For instance, the HCN~(1--0)/HCO$^{+}$(1--0) luminosity ratio of NGC~1068 
is estimated to be 1.64 $\pm$ 0.03 from single-dish measurements 
\citep[$\Omega_{\rm PB} \sim$ 24\arcsec;][]{2015A&A...579A.101A}.
Meanwhile, it is $\sim$2.0 at the AGN position and its surrounding regions 
using interferometric measurements 
\citep[$\theta_{\rm beam} \sim$ 6\arcsec;][]{2014A&A...570A..28V}.
Since the large beam of a single-dish covers various regions 
such as the nucleus, circumnuclear disk, and starburst ring of NGC~1068, 
the line ratio is affected by contaminations 
from those regions with different chemical compositions.

The HCN~(1--0)/HCO$^{+}$(1--0) luminosity ratios of the merger remnants 
are presented in Figure~\ref{fig:f4} (left).
The mean $L'_{\rm HCN}/L'_{\rm HCO^{+}}$ of the merger remnants is 0.88 $\pm$ 0.08.
The line ratios of two merger remnants that host X-ray detected AGNs are presented 
by the black open circles in Figure~\ref{fig:f4} (left).
Although they are higher than the mean ratio, 
they are similar to other merger remnants within the errors.
This result is consistent with the previous study by \citet{2015ApJ...814...39P}.
While they find that the HCN~(1--0) emission is enhanced in AGN-dominated systems, 
some composite and SB-dominated systems have HCN/HCO$^{+}$ line ratios 
which are comparable to those of AGN-dominated systems.
Therefore, the global low-$J$ HCN/HCO$^{+}$ line ratio should be treated 
with caution when it is used as an indicator of an AGN.

For comparison, we plot the HCN/HCO$^{+}$ luminosity ratios of 44 local U/LIRGs 
in the subsample of the GOALS \citep{2015ApJ...814...39P} in Figure~\ref{fig:f4} (left).
The mean $L'_{\rm HCN}/L'_{\rm HCO^{+}}$ of the late-stage mergers and early/mid-stage mergers 
are 1.02 $\pm$ 0.15 and 0.92 $\pm$ 0.12, respectively (Table~\ref{tab:t5}).
The mean ratio of the late-stage mergers decreases to 0.77 $\pm$ 0.08 
when the four AGN-dominated systems are excluded.
These mean ratios are consistent with that of the merger remnants.
There is no significant difference among the different merger stages at kpc scales.

\subsection{HNC~(1--0)/HCN~(1--0) ratio}
The HNC~(1--0)/HCN~(1--0) luminosity ratios of the merger remnants 
are presented in Figure~\ref{fig:f4} (right).
The HNC/HCN luminosity ratios of seven merger remnants 
detected in both molecular lines range from 0.31 to 0.70, 
and the mean line ratio is 0.43 $\pm$ 0.05.
The upper limits of $L'_{\rm HNC}/L'_{\rm HCN}$ of nine merger remnants are below 0.5.
These results are consistent with the previous measurements of local IR-bright galaxies 
\citep{2020A&A...633A.163C} and spiral galaxies \citep{2019ApJ...880..127J}, 
and suggest PDR dominated sources \citep{2008A&A...477..747B}.
We plot the HNC/HCN luminosity ratios of 44 local U/LIRGs 
\citep{2015ApJ...814...39P} in Figure~\ref{fig:f4} (right).
The HNC/HCN luminosity ratios of all the U/LIRGs except for one galaxy are lower than unity.
The mean $L'_{\rm HNC}/L'_{\rm HCN}$ of the late-stage mergers and early/mid-stage mergers 
are 0.52 $\pm$ 0.03 and 0.60 $\pm$ 0.09, respectively (Table~\ref{tab:t5}).
These are consistent with the average $L'_{\rm HNC}/L'_{\rm HCN}$ 
of the merger remnants within the errors.
Both HNC/HCN and HCN/HCO$^{+}$ line ratios do not show differences 
among the different merger stages in kpc scales.

The HNC/HCN line ratio has been proposed as a tool 
for probing the gas kinematic temperature ($T_{\rm kin}$), 
based on observations towards the Galactic sources.
The HNC/HCN abundance ratios are close to unity in dark cloud cores 
at lower temperatures \citep{1998ApJ...503..717H}, 
whereas they are as low as $\sim$0.013 
in the vicinity of the Orion KL hot cores \citep{1992A&A...256..595S}.
The HNC/HCN abundance ratio seems to decrease with increasing temperature. 
This is supported by the chemical models which predict that 
HNC tends to be selectively destroyed by neutral-neutral reactions 
at higher temperatures \citep{2014ApJ...787...74G}.
According to the empirical calibration derived by \citet{2020A&A...635A...4H}, 
the HNC/HCN luminosity ratios of 0.31--0.70 correspond to $T_{\rm kin}$ = 14--32~K, 
which are lower than the typical $T_{\rm kin}$ of star-forming galaxies 
estimated from ammonia observations \citep{2013ApJ...779...33M}.
Moreover, the HNC/HCN luminosity ratio of one non-merging LIRG exceeds unity, 
which corresponds to $T_{\rm kin} <$ 10~K.
Such low kinetic temperatures are unusual for IR-bright galaxies 
\citep[$T_{\rm kin} \geq$ 40~K;][]{2007A&A...464..193A}.
High HNC/HCN line ratios ($>$1) have been found in local U/LIRGs 
and could be explained by IR radiative pumping or by XDRs \citep{2007A&A...464..193A}.
From these observational results, we conclude that it is unlikely that 
the HNC/HCN line ratio averaged across an extragalactic source 
can be used as a tracer of the gas kinetic temperature.

\section{Dense gas fraction and star formation efficiency}
\subsection{HCN~(1--0)/CO~(1--0) ratio}
The HCN~(1--0)/CO~(1--0) line ratio has been used as a proxy 
for the dense gas fraction \citep{2004ApJS..152...63G, 2009ApJ...707.1217J}. 
We calculate the HCN~(1--0)/CO~(1--0) luminosity ratios ($L'_{\rm HCN}/L'_{\rm CO}$) 
of 20 merger remnants when at least either HCN or CO luminosity was determined. 
We used literature values of $L'_{\rm CO}$ 
measured with the IRAM 30~m telescope for 11/20 sources 
\citep{1997ApJ...478..144S, 1999AJ....118..145Z, 2005ApJ...624..714G, 
2006A&A...448...29B, 2010A&A...509A..19J, 2011A&A...528A..30C, 
2012A&A...539A...8G, 2015A&A...579A.101A, 2019A&A...628A..71H}. 
The beam of the IRAM 30~m telescope at $\sim$115~GHz ($\Omega_{\rm PB} \sim 22\arcsec$) 
is $\sim$2\arcsec smaller than the LMT beam at $\sim$88~GHz ($\Omega_{\rm PB} \sim 24\arcsec$).
To account for the unmatched beam sizes, 
we added an additional 20\% uncertainty to the CO luminosity 
when calculating the HCN/CO luminosity ratio. 
We made the same correction to the CO luminosity for the control samples. 
We used literature $L'_{\rm CO}$ of UGC~2238 measured with 
the National Radio Astronomy Observatory (NRAO) 12~m telescope \citep{1991ApJ...370..158S}.
The beam of the NRAO 12~m telescope ($\Omega_{\rm PB} \sim$ 55\arcsec) 
is much larger than the LMT beam, and the past CO imaging study reveals 
that the CO emission is distributed inside and outside the LMT beam 
\citep[$R_{\rm CO}$ = 20\farcs5;][]{2014ApJS..214....1U}.
Hence, we consider $L'_{\rm HCN}/L'_{\rm CO}$ of UGC 2238 as a lower limit. 
We calculated $L'_{\rm CO}$ of eight sources using the ALMA CO~(1--0) maps. 
We measured the CO integrated intensity within 24\arcsec, 
which is the same as the LMT beam at $\sim$88~GHz. 
We added an additional 10\% uncertainty to $L'_{\rm CO}$ of NGC~3256 
to account for the missing flux. 
Since the data of the remaining seven sources were corrected 
with the zero-spacing information, we have not made any additional correction. 
Similarly, we measured $L'_{\rm CO}$ of NGC~2782 
using the PdBI CO~(1--0) map \citep{2008A&A...482..133H}. 
\citet{2014ApJS..214....1U} have found that the data suffer from missing flux. 
The recovered flux is 54\%. We thus use $L'_{\rm CO}$ of NGC~2782 as the lower limit.
The literature and derived CO luminosities are summarized in Table~\ref{tab:t4}.

The HCN/CO luminosity ratios of the 20 merger remnants are shown 
as a function of the IR luminosity in Figure~\ref{fig:f5}.
There is a positive correlation between $L'_{\rm HCN}/L'_{\rm CO}$ and $L_{\rm IR}$.
In our sample, the average $L'_{\rm HCN}/L'_{\rm CO}$ of the U/LIRGs is 
three times higher than that of the non-LIRGs. 
This suggests that the U/LIRGs have a higher dense gas fraction 
when compared to the non-LIRGs.
The median $L'_{\rm HCN}/L'_{\rm CO}$ of the merger remnants is 0.045 (Table~\ref{tab:t6}).
Adapting the standard luminosity-to-mass conversion factors 
\citep[$\alpha_{\rm CO} = 4.35\,M_{\sun}$ (K~km~s$^{-1}$ pc$^{2}$)$^{-1}$ 
and $\alpha_{\rm HCN} = 10\,M_{\sun}$ (K~km~s$^{-1}$ pc$^{2}$)$^{-1}$;][]
{2013ARA&A..51..207B, 2004ApJS..152...63G}, 
the HCN/CO luminosity ratio of 0.041 corresponds 
to a dense gas fraction ($M_{\rm dense}$/$M_{\rm H_{2}}$) of $\sim$10\%, 
where $M_{\rm dense}$ is the dense molecular gas mass 
and $M_{\rm H_{2}}$ is the molecular gas mass. 
The largest HCN/CO luminosity ratios correspond to a dense gas fraction of order unity. 
It is known that the CO luminosity-to-mass conversion factor 
in U/LIRGs is lower than the standard value 
\citep[e.g., $\alpha_{\rm CO}$ = 0.6--0.8;][]{1998ApJ...507..615D, 2012ApJ...751...10P}.
When a small $\alpha_{\rm CO}$ is adopted, the dense gas fraction increases.

For comparison, we plot the HCN/CO luminosity ratios of 
the late-stage mergers, early/mid-stage mergers, and non-merging galaxies 
\citep{2015ApJ...814...39P, 2019A&A...628A..71H} and the LTGs \citep{2004ApJS..152...63G} 
in Figure~\ref{fig:f5}.
There appears to be little correlation between $L'_{\rm HCN}/L'_{\rm CO}$ and $L_{\rm IR}$ 
for $L_{\rm IR} < 10^{11}~L_{\sun}$, 
but $L'_{\rm HCN}/L'_{\rm CO}$ is strongly correlated with $L_{\rm IR}$ at higher $L_{\rm IR}$.
The histogram of $L'_{\rm HCN}/L'_{\rm CO}$ in each sample is shown in Figure~\ref{fig:f6} (left).
We compute the Kolmogorov-Smirnov (K-S) statistic 
on the merger remnant sample and the control sample.
The $P$-values from the K-S test are summarized in Table~\ref{tab:t7}.
The $P$-values for our sample and the samples in the different stages of mergers 
are large ($\geq$0.38), suggesting that the populations are not significantly different.
Numerical simulations predict that 
the dense gas fraction largely increases during the merging process, 
modifying the global gas density distribution of merging galaxies 
\citep[e.g.,][]{2009ApJ...707.1217J, 2011EAS....51..107B, 2019MNRAS.485.1320M}.
However, we do not find such a trend in this study, 
assuming that the same luminosity-to-mass conversion factors can be adopted
for all the mergers regardless of the merger stages. 
The mean $L'_{\rm HCN}/L'_{\rm CO}$ is consistent between 
the early/mid-stage and the late-stage (Table~\ref{tab:t6}), 
and decreases in the post-merger stage.
The HCN/CO luminosity ratios of the non-merging LIRGs 
are relatively high (Figure~\ref{fig:f6}), and 
the $P$-value for the merger remnants and the non-merging LIRGs is small ($\ll$0.05).
This implies that high dense gas fraction may play a crucial role 
in enhancing the star formation in non-merging galaxies, compared to merger remnants.

\subsection{IR-to-HCN~(1--0) Luminosity Ratio} 
We calculate the IR-to-HCN~(1--0) luminosity ratio, 
which is a proxy for the SFE$_{\rm dense}$ (= SFR/$M_{\rm dense}$). 
The IR/HCN luminosity ratios of the 21 merger remnants are shown 
as a function of the IR luminosity in Figure~\ref{fig:f7}. 
While the luminosity ratios vary from source to source 
(1.1 -- 6.5) $\times 10^{3}$ $L_{\sun}$/(K\,km\,s$^{-1}$\,pc$^{2}$)), 
there is no dependency of $L_{\rm IR}/L'_{\rm HCN}$ on $L_{\rm IR}$. 
The best-fit line for the merger remnants (the solid black line) 
is almost flat in Figure~\ref{fig:f7}. 
This can be expected based on the well-known correlation 
between $L'_{\rm HCN}$ and $L_{\rm IR}$ \citep{2004ApJ...606..271G}. 
However, the mean $L_{\rm IR}/L'_{\rm HCN}$ of the merger remnants 
is about four times higher than the mean $L_{\rm IR}/L'_{\rm HCN}$ 
($\sim$776 $L_{\sun}$/(K\,km\,s$^{-1}$\,pc$^{2}$)) 
derived using more than 800 data points of various sources, 
including resolved cores and Giant Molecular Clouds (GMCs) in the Galaxy, 
resolved galaxy disks, and entire galaxies \citep{2019ApJ...880..127J}.

The IR/HCN luminosity ratios of the control samples 
are also plotted in Figure~\ref{fig:f7}.
There appears to be a difference in $L_{\rm IR}/L'_{\rm HCN}$ among the samples.
The mean $L_{\rm IR}/L'_{\rm HCN}$ are 
$(2.9 \pm 0.5) \times 10^{3}$ for the 16 merger remnants with HCN detections, 
$(2.0 \pm 0.3) \times 10^{3}$ for the late-stage mergers, 
$(1.6 \pm 0.5) \times 10^{3}$ for the early/mid-stage mergers,
$(8.5 \pm 0.6) \times 10^{2}$ for the non-mering LIRGs, 
and $(7.5 \pm 1.0) \times 10^{2}$ for the LTGs 
in the unit of $L_{\sun}$/(K\,km\,s$^{-1}$\,pc$^{2}$).
We find that the mean ratio increases from the early-stage mergers 
to the late-stage mergers and the merger remnants (Table~\ref{tab:t6}). 
The mean $L_{\rm IR}/L'_{\rm HCN}$ of the merger remnants is more than 
three times higher than those of the non-merging LIRGs and the LTEs.
The histogram of $L_{\rm IR}/L'_{\rm HCN}$ in each sample 
is shown in Figure~\ref{fig:f6} (middle).
The K-S test gave $P$-value = 0.53 for our sample 
and the sample of late-stage mergers (Table~\ref{tab:t7}).
In contrast, it gave $P$-values smaller than 0.05 
for our sample and the other three control samples, 
suggesting that the populations are different.
According to \citet{2012A&A...539A...8G}, the SFE$_{\rm dense}$ averaged across the entire galaxy 
varies for different galaxy types.
They assemble a sample of $\sim$100 normal galaxies and U/LIRGs 
and find that $L_{\rm IR}/L'_{\rm HCN}$ is on average a few times higher 
in the U/LIRGs ($\sim$1400 $\pm$ 100) compared to the normal galaxies ($\sim$600 $\pm$ 70).
However, in the merger remnant sample, 
not only U/LIRGs but also non-LIRGs show high $L_{\rm IR}/L'_{\rm HCN}$.
Along with our new finding that the IR/HCN luminosity ratio increases along the merger sequence, 
these results suggest that the SFE$_{\rm dense}$ is enhanced during dynamical interactions and mergers.

\subsection{IR-to-CO~(1--0) Luminosity Ratio}
We calculate the IR-to-CO~(1--0) luminosity ratio, which is also a proxy for the SFE. 
The IR/CO luminosity ratios of the merger remnants and the control samples 
are plotted in Figure~\ref{fig:f8}. 
The luminosity ratios are distributed over two orders of magnitudes and increase with the IR luminosity, 
suggesting that the SFE is enhanced in active star-forming galaxies.
The mean $L_{\rm IR}/L'_{\rm CO}$ of the 18 merger remnants with CO detections 
is $(1.5 \pm 0.3) \times 10^2$ $L_{\sun}$/(K\,km\,s$^{-1}$\,pc$^{2}$). 
This is comparable to the late-stage mergers 
($(1.7 \pm 0.4) \times 10^{2}$ $L_{\sun}$/(K\,km\,s$^{-1}$\,pc$^{2}$)) 
and about two times higher than those of the early/mid-stage mergers and the non-merging LIRGs. 
These similarities and differences are also shown 
by the histograms of $L'_{\rm IR}/L_{\rm CO}$ (Figure~\ref{fig:f6} (right)) 
and the results of the K-S test (Table~\ref{tab:t7}). 
Overall, these results agree with the previous studies 
utilizing large samples of optically-selected galaxies. 
As \citet{2018MNRAS.476.2591V} find using the depletion time (= 1/SFE), 
the depletion time of the early-stage mergers, which are observed as galaxy pairs, 
are consistent with those of non-merging LIRGs. 
\citet{2018ApJ...868..132P} find that the SFE is enhanced 
in closer pairs of equal-mass galaxies. 
We also find a similar trend that the SFE is enhanced 
in the merger remnants and the late-stage mergers, 
although the exact fraction of these samples that result from a major merger is unknown 
since it is difficult to reverse the chronology 
and disentangle the exact mass and morphology of the progenitors.

\subsection{High star formation efficiency of merger remnants}
We find that, on average, the IR/HCN and IR/CO luminosity ratios of the merger remnants 
are comparable to those of the late-stage mergers 
and higher compared to the early/mid-stage mergers, the non-merging LIRGs, and the LTGs.
This result suggests that not only the SFE but also the SFE$_{\rm dense}$ are increased 
by dynamical interactions and mergers.
In addition, the SFEs do not change significantly between the pre- and post-coalescence.
A high efficiency of converting gas into stars has been commonly proposed 
as a possibility for intense star formation in galaxy mergers.
Recently observational studies have been conducted to check this enhancement 
\citep[e.g.,][]{2016PASJ...68...96M, 2018MNRAS.476.2591V, 2018ApJ...868..132P}.
As seen in the previous sections, 
we investigate the SFE$_{\rm dense}$ and SFE of galaxies at different stages of mergers, 
finding enhanced SFE$_{\rm dense}$ in the merger remnants.
The SFE enhancement in galaxy pairs and mergers has been predicted 
by simulation \citep[e.g.,][]{2007A&A...468...61D, 2020MNRAS.tmp.2894M}.
The parsec-resolution simulation shows that the compressive turbulence 
generates an excess of dense gas along the merging process, 
leading to an enhanced SFE \citep{2014MNRAS.442L..33R}.
By using zoom-in simulations of major mergers 
identified in the cosmological simulation, 
\citet{2016MNRAS.462.2418S} find that the SFR enhancement is driven 
by the increase in the molecular gas reservoir during the galaxy-pair period, 
leading to an increased SFE close to the final coalescence.
When the gas is compressed by dynamical interactions and reaches high densities, 
the star formation timescale is shorter than the cloud evaporation timescale,  
implying that a large fraction of the gas is converted into stars 
before the cold clouds are evaporated by thermal conduction. 
Therefore, stars form much more efficiently than normal star-forming gas.
This zoom-in simulation also shows that 
the gas is turned into stars with high efficiency 
even after the final coalescence \citep{2016MNRAS.462.2418S}.
Our result is consistent with this theoretical prediction, 
indicating that the merging process affects the star formation properties of the gas 
in the post-merger phase.

We re-compare $L_{\rm IR}/L'_{\rm HCN}$ between the merger remnants and the non-merging LIRGs 
by matching the range of their $L_{\rm IR}$. 
We select sources with $10^{11}\,(L_{\odot}) \leq L_{\rm IR} < 10^{12}\,(L_{\odot})$ 
(i.e., LIRGs) from the merger remnant sample and plot $L_{\rm IR}/L'_{\rm HCN}$
as a function of $L'_{\rm HCN}/L'_{\rm CO}$ in Figure~\ref{fig:f9}.
We find that these two samples occupy two different regions in the plot.
The merger remnants show higher SFE$_{\rm dense}$ and lower dense gas fraction, 
whereas the non-merging LIRGs show lower SFE$_{\rm dense}$ and higher dense gas fraction.
This suggests that the different modes of star formation may be taking place in these samples.
Merger remnants have a high SFE$_{\rm dense}$ due to the ISM turbulence \citep{2005ApJ...630..250K}
which could help compress the diffuse gas reservoirs,
hence efficiently boosting star formation.
On the other hand, star formation could be enhanced in non-merging galaxies 
by increasing the dense gas.
An enrichment of the molecular gas reservoir has also been proposed 
as a possibility for intense star formation.

Another possibility for the different distributions presented in Figure~\ref{fig:f9} 
is that the conversion factor from the HCN luminosity into the dense gas mass ($\alpha_{\rm HCN}$) 
is not the same between the merger remnants and the non-merging LIRGs. 
If $\alpha_{\rm HCN}$ for the merger remnants is higher by a factor of $\sim$3 
than $\alpha_{\rm HCN}$ for the non-merging LIRGs, 
the data points of the merger remnants would shift towards the bottom-right in Figure~\ref{fig:f9} 
and overlap with the distribution of the non-merging LIRGs. 
The dense gas conversion factor derived by \citet{2004ApJS..152...63G}
has been widely used for all extragalactic environments. 
This conversion factor is derived under the assumption that the HCN~(1--0) emission 
originates from a virialized cloud core with an average density of $3 \times 10^{4}$~cm$^{-3}$ 
and the brightness temperature of 35~K. 
However, the dense gas conversion factor depends on various factors 
such as the density and the excitation condition. 
If the brightness temperature of the HCN emitting clumps is larger than 35~K, 
or if their averaged density is less than $3 \times 10^{4}$ cm$^{-3}$, 
the conversion factor can become smaller \citep{2017A&A...604A..74S}.
Several observational studies and simulations have investigated variations in $\alpha_{\rm HCN}$
\citep[e.g.,][]{2007ApJ...654..304K, 2010ApJS..188..313W, 2017ApJ...835...76M, 2017A&A...602A..51V}.
\citet{2017A&A...604A..74S} compare the HCN~(1--0) luminosity in the Galactic GMCs 
with the dense molecular gas mass estimated from Herschel column density maps 
and suggest that variations in $\alpha_{\rm HCN}$ could be related to the far-ultraviolet field, 
that is, to gas excitation and/or chemistry variations. 
\citet{2018MNRAS.479.1702O} find that the dense gas conversion factor is $\alpha_{\rm HCN} \simeq 14$
$M_{\sun}$ (K~km~s$^{-1}$ pc$^{2}$)$^{-1}$ using hydrodynamical simulations, 
but the uncertainty is a factor of $\sim$2 mainly due to variations in the HCN abundance.
The dense gas conversion factor can vary from source to source, 
and the variation at a factor of $\sim$3 level could be plausible.

Another possibility is that the dense gas is rapidly depleted in the merger remnants, 
resulting in an apparent high SFE$_{\rm dense}$ and a lower dense gas fraction. 
\citet{2007A&A...468...61D} find that an increase in the SFR is correlated 
with an increase in the SFE in modeled galaxy pairs
and mergers, but the SFE peak is slightly delayed in time 
compared to the SFR peak for the most intense starbursts. 
The gas is more rapidly depleted by the merger-induced intense star formation, 
and then the amount of gas content decreases in merged galaxies. 
This condition can make an apparent high SFE as the SFE$_{\rm dense}$ is computed by SFR/$M_{\rm dense}$.

\section{Summary}
We have conducted multi-line observations towards 28 merger remnants using the LMT. 
Twenty-one out of 28 sources were detected in at least one molecular line. 
Seventeen sources were detected in either the HCN~(1--0) or the HCO$^{+}$(1--0) lines, 
and 20 sources were detected in the $^{13}$CO~(1--0) line.

Firstly, we calculated the intensity ratios of the dense gas tracers 
(HCN~(1--0), HCO$^{+}$(1--0), and HNC~(1--0)) to $^{13}$CO~(1--0). 
Then we divided our sample into two groups categorized by their IR luminosities 
(U/LIRGs and non-LIRGs) and compared the line ratios between the two groups.
There are significant differences in the line ratios. In particular, 
the HCN~(1--0)/$^{13}$CO~(1--0) ratios are different between the U/LIRG and non-LIRG groups. 
This difference is likely caused by the deficiency of $^{13}$CO, 
the IR radiative pumping, and high dense gas fractions in the U/LIRGs.

On the other hand, we do not find any significant differences 
in the line ratios between the dense gas tracers 
between the U/LIRGs and the non-LIRGs in our sample. 
The HNC~(1--0)/HCN~(1--0) luminosity ratios of the merger remnants are all below 0.70, 
suggesting PDR dominated sources. 
The HCN~(1--0)/HCO$^{+}$(1--0) luminosity ratios of two merger remnants 
which host X-ray detected AGNs are 
consistent with other merger remnants within the errors. 
As demonstrated by previous studies, 
the global HCN/HCO$^{+}$ line ratio is a less effective indicator of an AGN 
because of contaminations from the surrounding regions.

We calculate the HCN~(1--0)/CO~(1--0), IR-to-CO~(1--0), and IR-to-HCN~(1--0) luminosity ratios 
of the merger remnants and compare them with those of the control samples, 
including mergers in the different stages of mergers, non-merging LIRGs, and LTEs.
There is a positive correlation between $L_{\rm HCN}$/$L_{\rm CO}$ and $L_{\rm IR}$, 
suggesting that IR-bright galaxies have high dense gas fractions compared to normal galaxies.
Contrary to the theoretical prediction, the measured $L_{\rm HCN}$/$L_{\rm CO}$, 
which is a proxy of the dense gas fraction, 
does not show an enhancement along the merger stages.
We find that, on average, the IR/HCN and IR/CO luminosity ratios of the merger remnants 
are comparable to those of the late-stage mergers 
and a few times higher than those of the early/mid-stage mergers and the non-merging LIRGs.
This result suggests that (1) not only the SFE but also the SFE$_{\rm dense}$ are increased 
by the mechanisms related to the dynamical interactions and mergers, such as the ISM turbulence, 
and that (2) these SFEs are enhanced even after the final coalescence, 
indicating that the merging process affects the star formation properties of the gas 
in the post-merger phase.
Furthermore, the comparison between the merger remnants and the non-merging suggests 
that high dense gas fraction may play a crucial role in enhancing the star formation 
in non-merging galaxies, while the SFEs are enhanced by the dynamical interactions in mergers, 
hence efficiently boosting star formation

\begin{acknowledgments}
We thank G. C. Privon for kindly providing 
the data of the IR luminosity of their sample.

We also thank the LMT staff members from 
the National Institute of Astrophysics, Optics and Electronics 
and the University of Massachusetts Amherst 
for making these observations possible.

This work was supported by JSPS KAKENHI Grant Number JP19K14769.
DI is supported by JSPS KAKENHI Grant Number JP18H03725.

This paper makes use of the following ALMA data: 
ADS/JAO.ALMA\#2011.0.00099.S, 
ADS/JAO.ALMA\#2016.2.00006.S, 
ADS/JAO.ALMA\#2016.2.00042.S,
ADS/JAO.ALMA\#2016.2.00094.S, 
ADS/JAO.ALMA\#2017.1.01003.S, 
ADS/JAO.ALMA\#2018.1.00223.S.
ALMA is a partnership of ESO (representing its member states), NSF (USA) and NINS (Japan), 
together with NRC (Canada), MOST and ASIAA (Taiwan), and KASI (Republic of Korea), 
in cooperation with the Republic of Chile. 
The Joint ALMA Observatory is operated by ESO, AUI/NRAO and NAOJ.

This research has made use of the NASA/IPAC Extragalactic Database (NED) 
which is operated by the Jet Propulsion Laboratory, California Institute of Technology, 
under contract with the National Aeronautics and Space Administration. 
\end{acknowledgments}

\vspace{5mm}
\facilities{LMT, ALMA}

\software{astropy \citep{2013A&A...558A..33A},
          CASA \citep{2007ASPC..376..127M},
          linmix \citep{2007ApJ...665.1489K}
          }

\begin{deluxetable*}{lrrrrrRRrCc}
\tablecaption{Merger Remnant Sub-sample\label{tab:t1}}
\tablewidth{0pt}
\tabletypesize{\scriptsize}
\tablehead{
\colhead{Source}&\colhead{R.A.}&\colhead{Decl.}&\colhead{$D_{\rm L}$}&\colhead{Scale}&\colhead{V$_{\rm sys}$}&\colhead{log~$L_{\rm FIR}$}&\colhead{log~$L_{\rm IR}$}&\colhead{$R_{\rm CO}$}&\colhead{$R_{\rm ratio}$}&\colhead{classification}\\
&\colhead{(J2000)}&\colhead{(J2000)}&\colhead{(Mpc)}&\colhead{(pc/1\arcsec)}&\colhead{(km~s$^{-1}$)}&\colhead{($L_{\sun}$)}&\colhead{($L_{\sun}$)}&\colhead{(arcsec)}&
}
\decimalcolnumbers
\startdata
UGC~6 & 00 03 09 & 21 57 37 & 91.5 & 425 & 6582 & 10.91 & 11.05 & 2.94 & 1.1 \pm 0.4 & LIRG\\
NGC~34 & 00 11 06 & -12 06 26 & 82.6 & 385 & 5931 & 11.41 & 11.47 & 8.06 & 4.5 \pm 0.9 & LIRG\\ 
Arp~230  & 00 46 24 & -13 26 32 & 23.9 & 115 & 1742 & 9.57 & 9.64 & 27.7 & 2.6 \pm 0.2 & \nodata\\
NGC~828 & 02 10 09 & 39 11 25 & 74.6 & 349 & 5374 & 11.29 & 11.34 & 22.6 & 1.3 \pm 0.1 & LIRG\\ 
UGC~2238 & 02 46 17 & 13 05 44 & 89.8 & 417 & 6436 & 11.28 & 11.32 & 20.5 & 1.2 \pm 0.2 & LIRG\\
NGC~1614 & 04 33 59 & -08 34 44 & 66.1 & 311 & 4778 & 11.51 & 11.64 & 14.9 & 0.40 \pm 0.08 & LIRG\\
Arp~187 & 05 04 53 & -10 14 51 & 173.8 & 778 & 12291 & 10.80 & 10.88 & 11.9 & 0.50 \pm 0.05 & \nodata\\
AM~0612-373 & 06 13 47 & -37 40 37 & 136.5 & 621 & 9734 & <10.25 & <10.74 & 8.08 & 0.19 \pm 0.04 & \nodata\\
NGC~2623 & 08 38 24 & 25 45 17 & 77.1 & 360 & 5535 & 11.53 & 11.54 & 3.03 & 0.18 \pm 0.04 & LIRG\\
NGC~2782 & 09 14 05 & 40 06 49 & 35.1 & 168 & 2562 & 10.44 & 10.54 & 13.8 & 0.18 \pm 0.01 & \nodata\\
UGC~5101 & 09 35 51 & 61 21 11 & 166.8 & 749 & 11809 & 11.95 & 11.97 & 4.05 & 4.0 \pm 0.8 & LIRG\\
AM~0956-282 & 09 58 46 & -28 37 19 & 13.6 & 65 & 980 & <9.12 & <9.23 & 24.9 & 0.18 \pm 0.03 & \nodata\\
NGC~3256 & 10 27 51 & -43 54 13 & 37.6 & 179 & 2738 & 11.51 & 11.59 & 36.9 & 1.7 \pm 0.2 & LIRG\\
NGC~3597 & 11 14 42 & -23 43 40 & 48.5 & 230 & 3504 & 10.88 & 10.97 & 9.86 & 1.2 \pm 0.1 & \nodata\\
AM~1158-333 & 12 01 20 & -33 52 36 & 41.8 & 199 & 3027 & 9.96 & 10.04 & 5.25 & 0.48 \pm 0.04 & \nodata\\ 
NGC~4194 & 12 14 09 & 54 31 37 & 34.7 & 166 & 2506 & 10.81 & 10.91 & 13.1 & \nodata & \nodata\\
NGC~4441 & 12 27 20 & 64 48 05 & 36.8 & 175 & 2674 & 9.96 & 10.05 & 12.5 & 0.24 \pm 0.03 & \nodata\\
UGC~8058 & 12 56 14 & 56 52 25 & 178.6 & 797 & 12642 & 12.37 & 12.52 & 1.85 & 16 \pm 3 & ULIRG\\
AM~1255-430 & 12 58 08 & -43 19 47 & 126.6 & 578 & 9026 & <10.18 & <10.67 & 6.48 & 0.66 \pm 0.07 & \nodata\\
AM~1300-233 & 13 02 52 & -23 55 18 & 89.8 & 417 & 6446 & 11.41 & 11.41 & 5.78 & 0.08 \pm 0.01 & LIRG\\
Arp~193 & 13 20 35 & 34 08 22 & 97.5 & 451 & 7000 & 11.59 & 11.61 & 5.78 & 0.40 \pm 0.08 & LIRG\\
UGC~9829 & 15 23 01 & -01 20 50 & 118.8 & 545 & 8492 & 10.39 & 10.47 & 9.35 & \nodata & \nodata\\ 
NGC~6052 & 16 05 13 & 20 32 32 & 65.3 & 307 & 4716 & 10.88 & 10.94 & 12.4 & \nodata & \nodata\\
UGC~10675 & 17 03 15 & 31 27 29 & 142.5 & 646 & 10134 & 11.03 & 11.11 & 4.10 & 3 \pm 1 & LIRG\\
AM~2038-382 & 20 41 13 & -38 11 36 & 84.3 & 393 & 6057 & 10.37 & 10.52 & 5.53 & 1.2 \pm 0.1 & \nodata\\
AM~2055-425 & 20 58 26 & -42 39 00 & 181.7 & 810 & 12840 & 11.95 & 12.02 & 4.50 & 0.57 \pm 0.08 & ULIRG\\
NGC~7135 & 21 49 46 & -34 52 35 & 36.4 & 173 & 2640 & 9.05 & 9.13 & 2.47 & 0.01 \pm 0.01 & \nodata\\
NGC~7252 & 22 20 44 & -24 40 42 & 64.9 & 305 & 4688 & 10.66 & 10.73 & 8.51 & 0.60 \pm 0.05 & \nodata\\
\enddata
\tablecomments{
Column~1: source name. 
Column~2 and~3: right ascension and declination. 
Column~4: the luminosity distance. 
Column~5: the systemic velocity \citep{2004AJ....128.2098R}. 
Column~6 and~7: the FIR and IR luminosity estimated using the $IRAS$ catalogs, 
but the IR luminosities of Arp~187, AM~1158-333, UGC~9829, UGC~10675, and NGC~7135 
are estimated using $L_{\rm IR} = 1.2 \times L_{\rm FIR}$.
This is based on the average $L_{\rm IR}/L_{\rm FIR}$ (1.2 $\pm$ 0.1) 
of our sample except for two AGN-host galaxies.
Column~8: the radius enclosing the maximum extent of the CO distribution \citep{2014ApJS..214....1U}. 
Column~9: the extent of molecular gas relative to the stellar component \citep{2014ApJS..214....1U}. 
See the details in the text.
}
\end{deluxetable*}

\begin{deluxetable*}{lcccC}
\tablecaption{ALMA CO~(1--0) Maps\label{tab:t2}}
\tablewidth{0pt}
\tabletypesize{\scriptsize}
\tablehead{
\colhead{Source}&\colhead{Array}&\colhead{Velocity Res.}&\colhead{rms}&\colhead{Beam Size}\\
&&\colhead{(km~s$^{-1}$)}&\colhead{(mJy beam$^{-1}$)}&\colhead{(arcsec)}
}
\decimalcolnumbers
\startdata
Arp~187 & 12M+7M+TP & 20 & 2.04 & 2.98 $\times$ 1.79\\
AM~0956-282 & 12M+7M+TP & 5 & 4.36 & 4.86 $\times$ 3.22\\
NGC~3256 & 12M+7M & 2.5 & 1.95 & 2.22 $\times$ 2.22\\
NGC~3597 & 12M+7M+TP & 20 & 3.57 & 1.99 $\times$ 1.39\\
AM~1300-233 & 12M+7M+TP & 20 & 1.86 & 2.35 $\times$ 1.20\\
AM~2055-425 & 12M+7M+TP & 20 & 2.15 & 1.53 $\times$ 1.21\\
NGC~7252 & 12M+7M+TP & 20 & 2.83 & 2.57 $\times$ 1.66\\
\enddata
\tablecomments{
Column~1: source name.
Column~2: array name (12M = the 12~m array, 7M = the 7~m array, TP = the total power array).
Column~3: the velocity resolution of the channel map.
Column~4: the noise level per channel.
Column~5: the synthesized beam size.
}
\end{deluxetable*}

\clearpage
\startlongtable
\begin{deluxetable*}{llR|RRR|RR}
\tablecaption{Properties of Molecular Lines\label{tab:t3}}
\tablewidth{0pt}
\tabletypesize{\scriptsize}
\tablehead{
\colhead{Source}&\colhead{Line}&\colhead{$V_{\rm LSR}$}&\colhead{$T_{\rm A, peak}^{*}$}&\colhead{$N_{\rm ch}$}&\colhead{$N_{\rm ch} \times \Delta\,V_{\rm ch}$}&\colhead{$I$}&\colhead{$S\Delta V$}\\
&&\colhead{(km s$^{-1}$)}&\colhead{(mK)}&&\colhead{(km s$^{-1}$)}&\colhead{(K($T_{\rm A}^{*}$)~km~s$^{-1}$)}&\colhead{(Jy~km~s$^{-1}$)}
 }
\decimalcolnumbers
\startdata
UGC~6 & HCN~(1--0) & \nodata & <1.20 & 4 & 423 & <0.25 & <1.8\\
($\sigma$ = 0.40 mK) & HCO$^{+}$(1--0) & \nodata & <1.20 & 4 & 420 & <0.25 & <1.8\\
 & HNC~(1--0) & \nodata & <1.20 & 4 & 413 & <0.25 & <1.7\\
 & CS~(2--1) & \nodata & <1.20 & 5 & 478 & <0.26 & <1.8\\
 & C$^{18}$O~(1--0) & \nodata & <1.20 & 5 & 427 & <0.23 & <1.6\\
 & $^{13}$CO~(1--0) & \nodata & <1.20 & 5 & 425 & <0.23 & <1.6\\
\hline
NGC~34 & HCN~(1--0) & 5747 \pm 32 & 1.43 \pm 0.33 & 5 & 529 & 0.62 \pm 0.08 & 4.3 \pm 0.5\\
($\sigma$ = 0.33 mK) & HCO$^{+}$(1--0) & 5716 \pm 26 & 1.50 \pm 0.33 & 4 & 420 & 0.54 \pm 0.07 & 3.8 \pm 0.5\\
 & HNC~(1--0) & \nodata & <0.99 & 5 & 517 & <0.23 & <1.6\\
 & CS~(2--1) & \nodata & <0.99 & 5 & 478 & <0.21 & <1.5\\
 & C$^{18}$O~(1--0) & \nodata & <0.99 & 4 & 341 & <0.17 & <1.2\\
 & $^{13}$CO~(1--0) & 5768 \pm 21 & 1.90 \pm 0.33 & 4 & 340 & 0.43 \pm 0.06 & 3.0 \pm 0.4\\
\hline
Arp~230 & HCN~(1--0) & \nodata & <0.81 & 3 & 317 & <0.15 & <1.0\\
($\sigma$ = 0.27 mK) & HCO$^{+}$(1--0) & \nodata & <0.81 & 3 & 315 & <0.15 & <1.0\\
 & HNC~(1--0) & \nodata & <0.81 & 3 & 310 & <0.14 & <1.0\\
 & CS~(2--1) & \nodata & <0.81 & 3 & 287 & <0.13 & <0.94\\
 & C$^{18}$O~(1--0) & \nodata & <0.81 & 3 & 256 & <0.12 & <0.84\\
 & $^{13}$CO~(1--0) & \nodata & <0.81 & 3 & 255 & <0.12 & <0.83\\
\hline
NGC~828 & C$_{2}$H~(1--0) & \nodata & 1.41 \pm 0.28 & 7 & 750 & 0.69 \pm 0.11 & 4.9 \pm 0.8\\
($\sigma$ = 0.38 mK) & HCN~(1--0) & 5209 \pm 18 & 3.52 \pm 0.38 & 6 & 634 & 1.32 \pm 0.10 & 9.2 \pm 0.7\\
 & HCO$^{+}$(1--0) & 5195 \pm 16 & 3.87 \pm 0.38 & 5 & 525 & 1.25 \pm 0.09 & 8.7 \pm 0.6\\
 & HNC~(1--0) & 5236 \pm 43 & 1.31 \pm 0.38 & 4 & 413 & 0.40 \pm 0.08 & 2.8 \pm 0.5\\
 & CS~(2--1) & \nodata & <1.14 & 5 & 478 & <0.24 & <1.7\\
 & C$^{18}$O~(1--0) & 5192 \pm 18 & 2.38 \pm 0.38 & 5 & 427 & 0.86 \pm 0.07 & 6.1 \pm 0.5\\
 & $^{13}$CO~(1--0) & 5211 \pm 8 & 8.44 \pm 0.38 & 7 & 595 & 3.38 \pm 0.09 & 23.7 \pm 0.6\\
\hline
UGC~2238 & HCN~(1--0) & 6323 \pm 60 & 1.42 \pm 0.38 & 5 & 529 & 0.56 \pm 0.09 & 3.9 \pm 0.6\\
($\sigma$ = 0.38 mK) & HCO$^{+}$(1--0) & 6347 \pm 56 & 1.38 \pm 0.38 & 4 & 420 & 0.48 \pm 0.08 & 3.3 \pm 0.6\\
 & HNC~(1--0) & \nodata & <1.14 & 5 & 517 & <0.26 & <1.8\\
 & CS~(2--1) & \nodata & <1.14 & 5 & 478 & <0.24 & <1.7\\
 & C$^{18}$O~(1--0) & \nodata & <1.14 & 6 & 512 & <0.24 & <1.7\\
 & $^{13}$CO~(1--0) & 6361 \pm 14 & 3.30 \pm 0.38 & 6 & 510 & 1.28 \pm 0.08 & 8.9 \pm 0.6\\
\hline
NGC~1614 & C$_{2}$H~(1--0) & \nodata & 1.68 \pm 0.32 & 10 & 1072 & 1.13 \pm 0.11 & 7.9 \pm 0.8\\
($\sigma$ = 0.32 mK) & HCN~(1--0) & 4690 \pm 16 & 2.87 \pm 0.32 & 4 & 423 & 0.86 \pm 0.07 & 6.0 \pm 0.5\\
 & HCO$^{+}$(1--0) & 4688 \pm 11 & 4.58 \pm 0.32 & 5 & 525 & 1.32 \pm 0.08 & 9.3 \pm 0.5\\
 & HNC~(1--0) & 4673 \pm 35 & 1.19 \pm 0.32 & 3 & 310 & 0.31 \pm 0.06 & 2.2 \pm 0.4\\
 & CS~(2--1) & 4690 \pm 36 & 1.35 \pm 0.32 & 3 & 287 & 0.30 \pm 0.05 & 2.1 \pm 0.4\\
 & C$^{18}$O~(1--0) & \nodata & <0.96 & 5 & 427 & <0.18 & <1.3\\
 & $^{13}$CO~(1--0) & 4707 \pm 11 & 3.54 \pm 0.32 & 5 & 425 & 0.86 \pm 0.06 & 6.0 \pm 0.4\\
\hline
Arp~187 & HCN~(1--0) & \nodata & <0.90 & 6 & 634 & <0.23 & <1.6\\
($\sigma$ = 0.30 mK) & HCO$^{+}$(1--0) & \nodata & <0.90 & 6 & 630 & <0.23 & <1.6\\
 & HNC~(1--0) & \nodata & <0.90 & 6 & 620 & <0.23 & <1.6\\
 & CS~(2--1) & \nodata & <0.90 & 7 & 669 & <0.23 & <1.6\\
 & C$^{18}$O~(1--0) & \nodata & <0.90 & 7 & 597 & <0.20 & <1.4\\
 & $^{13}$CO~(1--0) & 11467 \pm 32 & 1.29 \pm 0.30 & 7 & 595 & 0.62 \pm 0.07 & 4.3 \pm 0.5\\
\hline
AM~0612-373 & HCN~(1--0) & \nodata & <0.87 & 7 & 740 & <0.24 & <1.7\\
($\sigma$ = 0.29 mK) & HCO$^{+}$(1--0) & \nodata & <0.87 & 7 & 735 & <0.24 & <1.7\\
 & HNC~(1--0) & \nodata & <0.87 & 7 & 723 & <0.24 & <1.7\\
 & CS~(2--1) & \nodata & <0.87 & 8 & 765 & <0.24 & <1.6\\
 & C$^{18}$O~(1--0) & \nodata & <0.87 & 9 & 768 & <0.22 & <1.6\\
 & $^{13}$CO~(1--0) & \nodata & <0.87 & 9 & 765 & <0.22 & <1.6\\
\hline
NGC~2623 & C$_{2}$H~(1--0) & \nodata & 1.32 \pm 0.15 & 8 & 858 & 0.62 \pm 0.05 & 4.3 \pm 0.3\\
($\sigma$ = 0.15 mK) & HCN~(1--0) & 5420 \pm 12 & 2.26 \pm 0.15 & 5 & 529 & 0.77 \pm 0.04 & 5.4 \pm 0.2\\
 & HCO$^{+}$(1--0) & 5412 \pm 13 & 2.14 \pm 0.15 & 5 & 525 & 0.72 \pm 0.04 & 5.0 \pm 0.2\\
 & HNC~(1--0) & 5435 \pm 12 & 1.72 \pm 0.15 & 5 & 517 & 0.56 \pm 0.03 & 3.9 \pm 0.2\\
 & CS~(2--1) & 5371 \pm 26 & 0.70 \pm 0.15 & 3 & 287 & 0.18 \pm 0.02 & 1.3 \pm 0.2\\
 & C$^{18}$O~(1--0) & 5399 \pm 22 & 0.66 \pm 0.15 & 4 & 341 & 0.18 \pm 0.03 & 1.3 \pm 0.2\\
 & $^{13}$CO~(1--0) & 5434 \pm 12 & 1.43 \pm 0.15 & 5 & 425 & 0.47 \pm 0.03 & 3.3 \pm 0.2\\
\hline
NGC~2782 & HCN~(1--0) & 2583 \pm 36 & 1.32 \pm 0.27 & 5 & 529 & 0.45 \pm 0.06 & 3.2 \pm 0.4\\
($\sigma$ = 0.27 mK) & HCO$^{+}$(1--0) & 2570 \pm 18 & 2.37 \pm 0.27 & 4 & 420 & 0.65 \pm 0.06 & 4.6 \pm 0.4\\
 & HNC~(1--0) & \nodata & <0.81 & 4 & 413 & <0.17 & <1.2\\
 & CS~(2--1) & 2524 \pm 45 & 0.83 \pm 0.27 & 3 & 287 & 0.18 \pm 0.04 & 1.3 \pm 0.3\\
 & C$^{18}$O~(1--0) & 2488 \pm 40 & 0.93 \pm 0.27 & 5 & 427 & 0.25 \pm 0.05 & 1.8 \pm 0.4\\
 & $^{13}$CO~(1--0) & 2515 \pm 19 & 2.16 \pm 0.27 & 6 & 510 & 0.75 \pm 0.06 & 5.2 \pm 0.4\\
\hline
UGC~5101 & C$_{2}$H~(1--0) & \nodata & 0.83 \pm 0.14 & 9 & 965 & 0.44 \pm 0.05 & 3.1 \pm 0.3\\
($\sigma$ = 0.14 mK) & HCN~(1--0) & 11346 \pm 23 & 1.75 \pm 0.14 & 7 & 740 & 0.98 \pm 0.04 & 6.8 \pm 0.3\\
 & HCO$^{+}$(1--0) & 11332 \pm 32 & 1.17 \pm 0.14 & 9 & 945 & 0.75 \pm 0.04 & 5.3 \pm 0.3\\
 & HNC~(1--0) & 11321 \pm 32 & 0.89 \pm 0.14 & 7 & 723 & 0.51 \pm 0.04 & 3.6 \pm 0.3\\
 & HC$_{3}$N~(10--9) & 11328 \pm 69 & 0.45 \pm 0.14 & 6 & 618 & 0.19 \pm 0.04 & 1.3 \pm 0.2\\
 & CS~(2--1) & 11388 \pm 36 & 0.60 \pm 0.14 & 7 & 669 & 0.32 \pm 0.04 & 2.2 \pm 0.2\\
 & C$^{18}$O~(1--0) & 11350 \pm 41 & 0.51 \pm 0.14 & 6 & 512 & 0.20 \pm 0.03 & 1.4 \pm 0.2\\
 & $^{13}$CO~(1--0) & 11312 \pm 24 & 0.95 \pm 0.14 & 7 & 595 & 0.47 \pm 0.03 & 3.3 \pm 0.2\\
 & CN~(1--0; 1/2--1/2) & 11367 \pm 25 & 1.19 \pm 0.14 & 9 & 745 & 0.68 \pm 0.03 & 4.8 \pm 0.2\\
 & CN~(1--0; 3/2--1/2) & 11340 \pm 21 & 1.86 \pm 0.14 & 10 & 825 & 1.11 \pm 0.04 & 7.8 \pm 0.3\\
\hline
AM~0956-282 & HCN~(1--0) & \nodata & <0.90 & 4 & 423 & <0.19 & <1.3\\
($\sigma$ = 0.30 mK) & HCO$^{+}$(1--0) & \nodata & <0.90 & 4 & 420 & <0.19 & <1.3\\
 & HNC~(1--0) & \nodata & <0.90 & 4 & 413 & <0.19 & <1.3\\
 & CS~(2--1) & 1020 \pm 29 & 1.01 \pm 0.30 & 4 & 382 & 0.28 \pm 0.06 & 1.9 \pm 0.4\\
 & C$^{18}$O~(1--0) & \nodata & <0.90 & 3 & 256 & <0.13 & <0.93\\
 & $^{13}$CO~(1--0) & 1051 \pm 31 & $\sim$0.79\tablenotemark{a} & 3 & 255 & 0.18 \pm 0.04 & 1.2 \pm 0.3\\
\hline
NGC~3256 & c-C$_{3}$H$_{2}$~(2$_{1,2}$--1$_{1,0}$) & 2785 \pm 22 & 1.26 \pm 0.33 & 2 & 220 & 0.77 \pm 0.05 & 1.9 \pm 0.4\\
($\sigma$ = 0.33 mK) & C$_{2}$H~(1--0) & \nodata & 5.51 \pm 0.33 & 6 & 644 & 1.83 \pm 0.09 & 12.8 \pm 0.6\\
 & HCN~(1--0) & 2793 \pm 4 & 13.59 \pm 0.33 & 4 & 423 & 2.91 \pm 0.07 & 20.4 \pm 0.5\\
 & HCO$^{+}$(1--0) & 2794 \pm 5 & 19.21 \pm 0.33 & 5 & 525 & 4.74 \pm 0.08 & 33.2 \pm 0.5\\
 & HNC~(1--0) & 2797 \pm 8 & 6.33 \pm 0.33 & 3 & 310 & 1.23 \pm 0.06 & 8.6 \pm 0.4\\
 & CH$_{3}$OH~(2$_{k}$--1$_{k}$) & 2820 \pm28 & 1.31 \pm 0.33 & 3 & 322 & 0.32 \pm 0.06 & 2.2 \pm 0.4\\
 & CS~(2--1) & 2780 \pm 8 & 4.83 \pm 0.33 & 4 & 382 & 1.04 \pm 0.06 & 7.3 \pm 0.4\\
 & CH$_{3}$CCH~(6$_{k}$--5$_{k}$) & 2766 \pm 31 & 1.17 \pm 0.33 & 2 & 183 & 0.20 \pm 0.04 & 1.4 \pm 0.3\\
 & C$^{18}$O~(1--0) & 2780 \pm 8 & 3.54 \pm 0.33 & 4 & 341 & 0.74 \pm 0.06 & 5.2 \pm 0.4\\
 & $^{13}$CO~(1--0) & 2792 \pm 4 & 15.88 \pm 0.33 & 4 & 340 & 3.29 \pm 0.06 & 23.0 \pm 0.4\\
\hline
NGC~3597 & HCN~(1--0) & 3469 \pm 28 & 0.90 \pm 0.27 & 3 & 317 & 0.23 \pm 0.05 & 1.6 \pm 0.3\\
($\sigma$ = 0.27 mK) & HCO$^{+}$(1--0) & 3491 \pm 22 & 1.32 \pm 0.27 & 3 & 315 & 0.32 \pm0.05 & 2.2 \pm 0.3\\
 & HNC~(1--0) & \nodata & <0.81 & 3 & 310 & <0.14 & <1.0\\
 & CS~(2--1) & \nodata & <0.81 & 3 & 287 & <0.13 & <0.94\\
 & C$^{18}$O~(1--0) & \nodata & <0.81 & 4 & 341 & <0.14 & <0.97\\
 & $^{13}$CO~(1--0) & 3464 \pm 11 & 2.63 \pm 0.27 & 4 & 340 & 0.76 \pm 0.05 & 5.3 \pm 0.3\\
\hline
AM~1158-333 & HCN~(1--0) & \nodata & <1.05 & 2 & 211 & <0.16 & <1.1\\
($\sigma$ = 0.35 mK) & HCO$^{+}$(1--0) & \nodata & <1.05 & 2 & 210 & <0.16 & <1.1\\
 & HNC~(1--0) & \nodata & <1.05 & 2 & 207 & <0.15 & <1.1\\
 & CS~(2--1) & \nodata & <1.05 & 2 & 191 & <0.14 & <0.99\\
 & C$^{18}$O~(1--0) & \nodata & <1.05 & 3 & 256 & <0.16 & <1.1\\
 & $^{13}$CO~(1--0) & \nodata & <1.05 & 3 & 255 & <0.15 & <1.1\\
\hline
NGC~4194 & HCN~(1--0) & 2521 \pm 30 & 1.62 \pm 0.36 & 3 & 317 & 0.36 \pm 0.07 & 2.6 \pm 0.5\\
($\sigma$ = 0.36mK) & HCO$^{+}$(1--0) & 2542 \pm 11 & 3.31 \pm 0.36 & 3 & 315 & 0.70 \pm 0.07 & 4.9 \pm 0.5\\
 & HNC~(1--0) & \nodata & <1.08 & 3 & 310 & <0.19 & <1.4\\
 & CS~(2--1) & \nodata & <1.08 & 3 & 287 & <0.18 & <1.3\\
 & C$^{18}$O~(1--0) & \nodata & <1.08 & 3 & 256 & <0.16 & <1.1\\
 & $^{13}$CO~(1--0) & 2517 \pm 12 & 2.87 \pm 0.36 & 3 & 255 & 0.56 \pm 0.05 & 4.0 \pm 0.4\\
\hline
NGC~4441 & HCN~(1--0) & \nodata & <0.93 & 3 & 317 & <0.17 & <1.2\\
($\sigma$ = 0.31 mK) & HCO$^{+}$(1--0) & \nodata & <0.93 & 3 & 315 & <0.17 & <1.2\\
 & HNC~(1--0) & \nodata & <0.93 & 3 & 310 & <0.17 & <1.2\\
 & CS~(2--1) & \nodata & <0.93 & 3 & 287 & <0.15 & <1.1\\
 & C$^{18}$O~(1--0) & \nodata & <0.93 & 3 & 256 & <0.14 & <0.96\\
 & $^{13}$CO~(1--0) & 2691 \pm 28 & 1.29 \pm 0.31 & 3 & 255 & 0.27 \pm 0.05 & 1.9 \pm 0.3\\
\hline
UGC~8058 & C$_{2}$H~(1--0) & \nodata & 0.99 \pm 0.23 & 6 & 643 & 0.41 \pm 0.06 & 2.9 \pm 0.4\\
($\sigma$ = 0.23 mK) & HCN~(1--0) & 12123 \pm 12 & 4.18 \pm 0.23 & 9 & 951 & 1.67 \pm 0.07 & 11.7 \pm 0.5\\
 & HCO$^{+}$(1--0) & 12125 \pm 10 & 3.56 \pm 0.23 & 11 & 1155 & 1.56 \pm 0.08 & 10.9 \pm 0.6\\
 & HNC~(1--0) & 12134 \pm 12 & 2.41 \pm 0.23 & 4 & 413 & 0.59 \pm 0.05 & 4.1 \pm 0.3\\
 & HC$_{3}$N~(10--9) & 12147 \pm 28 & 0.89 \pm 0.23 & 2 & 206 & 0.16 \pm 0.03 & 1.1 \pm 0.2\\
 & N$_{2}$H$^{+}$(1--0) & 12163 \pm 34 & 0.69 \pm 0.23 & 2 & 201 & 0.12 \pm 0.03 & 0.8 \pm 0.2\\
 & CS~(2--1) & 12113 \pm 23 & 1.00 \pm 0.23 & 2 & 191 & 0.17 \pm 0.03 & 1.2 \pm 0.2\\
 & HC$_{3}$N~(12--11) & 12205 \pm 24 & 1.47 \pm 0.23 & 5 & 429 & 0.40 \pm 0.04 & 2.8 \pm 0.3\\
 & C$^{18}$O~(1--0) & \nodata & <0.69 & 2 & 171 & <0.08 & <0.58\\
 & $^{13}$CO~(1--0) & 12139 \pm 28 & 0.84 \pm 0.23 & 2 & 170 & 0.13 \pm 0.03 & 0.9 \pm 0.2\\
 & CN~(1--0; 1/2--1/2) & 12122 \pm 15 & 1.96 \pm 0.23 & 5 & 414 & 0.63 \pm 0.04 & 4.4 \pm 0.3\\
 & CN~(1--0; 3/2--1/2) & 12127 \pm 9 & 4.45 \pm 0.23 & 5 & 413 & 1.05 \pm 0.04 & 7.3 \pm 0.3\\
 & $^{12}$CO~(1--0) & 12138 \pm 1 & 36.35 \pm 0.23 & 17 & 1382 & 9.61 \pm 0.08 & 67.2 \pm 0.5\\
\hline
AM~1255-430 & HCN~(1--0) & \nodata & <0.87 & 5 & 529 & <0.21 & <1.4\\
($\sigma$ = 0.29 mK) & HCO$^{+}$(1--0) & \nodata & <0.87 & 5 & 525 & <0.20 & <1.4\\
 & HNC~(1--0) & \nodata & <0.87 & 5 & 517 & <0.20 & <1.4\\
 & CS~(2--1) & \nodata & <0.87 & 5 & 478 & <0.19 & <1.3\\
 & C$^{18}$O~(1--0) & \nodata & <0.87 & 6 & 512 & <0.18 & <1.3\\
 & $^{13}$CO~(1--0) & \nodata & <0.87 & 6 & 510 & <0.18 & <1.3\\
\hline
AM~1300-233 & HCN~(1--0) & 6308 \pm 58 & 0.84 \pm 0.24 & 7 & 740 & 0.40 \pm 0.07 & 2.8 \pm 0.5\\
($\sigma$ = 0.24 mK) & HCO$^{+}$(1--0) & 6313 \pm 29 & 1.72 \pm 0.24 & 6 & 630 & 0.76 \pm 0.06 & 5.3 \pm 0.4\\
 & HNC~(1--0) & \nodata & <0.72 & 6 & 620 & <0.18 & <1.3\\
 & CS~(2--1) & 6290 \pm 59 & $\sim$0.68\tablenotemark{a} & 6 & 574 & 0.30 \pm 0.06 & 2.1 \pm 0.4\\
 & C$^{18}$O~(1--0) & \nodata & <0.72 & 6 & 512 & <0.15 & <1.1\\
 & $^{13}$CO~(1--0) & 6269 \pm 37 & 1.33 \pm 0.24 & 6 & 510 & 0.41 \pm 0.05 & 2.8 \pm 0.3\\
 & CN~(1--0; 1/2--1/2) & 6299 \pm 49 & 0.85 \pm 0.24 & 6 & 497 & 0.29 \pm 0.05 & 2.1 \pm 0.3\\
\hline
Arp~193 & C$_{2}$H~(1--0) & 6762 \pm 39 & 1.49 \pm 0.29 & 9 & 965 & 0.75 \pm 0.09 & 5.2 \pm 0.7\\
($\sigma$ = 0.29 mK) & HCN~(1--0) & 6763 \pm 18 & 2.50 \pm 0.29 & 5 & 529 & 0.84 \pm 0.07 & 5.9 \pm 0.5\\
 & HCO$^{+}$(1--0) & 6800 \pm 15 & 3.42 \pm 0.29 & 5 & 525 & 1.29 \pm 0.07 & 9.1 \pm 0.5\\
 & HNC~(1--0) & 6809 \pm 52 & 0.92 \pm 0.29 & 5 & 517 & 0.41 \pm 0.07 & 2.9 \pm 0.5\\
 & CS~(2--1) & 6765 \pm 40 & 1.20 \pm 0.29 & 5 & 478 & 0.38 \pm 0.06 & 2.7 \pm 0.4\\
 & C$^{18}$O~(1--0) & \nodata & <0.87 & 5 & 427 & <0.17 & <1.2\\
 & $^{13}$CO~(1--0) & 6826 \pm 19 & 2.38 \pm 0.29 & 5 & 425 & 0.78 \pm 0.06 & 5.4 \pm 0.4\\
 & CN~(1--0; 1/2--1/2) & 6858 \pm 28 & 1.67 \pm 0.29 & 7 & 579 & 0.56 \pm 0.06 & 3.9 \pm 0.4\\
\hline
UGC~9829 & HCN~(1--0) & \nodata & <0.75 & 5 & 529 & <0.18 & <1.2\\
($\sigma$ = 0.25 mK) & HCO$^{+}$(1--0) & \nodata & <0.75 & 5 & 525 & <0.18 & <1.2\\
 & HNC~(1--0) & \nodata & <0.75 & 5 & 517 & <0.17 & <1.2\\
 & CS~(2--1) & \nodata & <0.75 & 5 & 478 & <0.16 & <1.1\\
 & C$^{18}$O~(1--0) & \nodata & <0.75 & 5 & 427 & <0.14 & <1.0\\
 & $^{13}$CO~(1--0) & 8290 \pm 22 & 1.63 \pm 0.25 & 5 & 425 & 0.43 \pm 0.05 & 3.0 \pm 0.3\\
\hline
NGC~6052 & C$_{2}$H~(1--0) & 4659 \pm 25 & 1.18 \pm 0.31 & 2 & 215 & 0.23 \pm 0.05 & 1.6 \pm 0.3\\
($\sigma$ = 0.31 mK) & HCN~(1--0) & \nodata & <0.93 & 3 & 317 & <0.17 & <1.2\\
 & HCO$^{+}$(1--0) & 4636 \pm 17 & 1.96 \pm 0.31 & 3 & 315 & 0.37 \pm 0.06 & 2.6 \pm 0.4\\
 & HNC~(1--0) & \nodata & <0.93 & 3 & 310 & <0.17 & <1.2\\
 & CS~(2--1) & \nodata & <0.93 & 3 & 287 & <0.15 & <1.1\\
 & C$^{18}$O~(1--0) & \nodata & <0.93 & 3 & 256 & <0.14 & <0.96\\
 & $^{13}$CO~(1--0) & 4617 \pm 13 & 2.05 \pm 0.31 & 3 & 255 & 0.34 \pm 0.05 & 2.4 \pm 0.3\\
\hline
UGC~10675 & HCN~(1--0) & 9808 \pm 24 & 1.34 \pm 0.27 & 3 & 317 & 0.19 \pm 0.05 & 1.3 \pm 0.3\\ 
($\sigma$ = 0.27 mK) & HCO$^{+}$(1--0) & \nodata & <0.81 & 3 & 315 & <0.15 & <1.0\\ 
 & HNC~(1--0) & \nodata & <0.81 & 3 & 310 & <0.14 & <1.0\\ 
 & CS~(2--1) & \nodata & <0.81 & 3 & 287 & <0.13 & <0.94\\ 
 & C$^{18}$O~(1--0) & \nodata & <0.81 & 4 & 341 & <0.14 & <0.97\\
 & $^{13}$CO~(1--0) & \nodata & <0.81 & 4 & 340 & <0.14 & <0.96\\ 
 & CN~(1--0; 1/2--1/2) & 9817 \pm 26 & 0.92 \pm 0.27 & 2 & 166 & 0.11 \pm 0.03 & 0.8 \pm 0.2\\ 
 & CN~(1--0; 3/2--1/2) & 9895 \pm 43 & 0.96 \pm 0.27 & 6 & 495 & 0.34 \pm 0.05 & 2.4 \pm 0.4\\
\hline
AM~2038-382 & HCN~(1--0) & \nodata & <0.93 & 5 & 529 & <0.22 & <1.5\\
($\sigma$ = 0.31 mK) & HCO$^{+}$(1--0) & \nodata & <0.93 & 5 & 525 & <0.22 & <1.5\\
 & HNC~(1--0) & \nodata & <0.93 & 5 & 517 & <0.21 & <1.5\\
 & CS~(2--1) & \nodata & <0.93 & 5 & 478 & <0.20 & <1.4\\
 & C$^{18}$O~(1--0) & \nodata & <0.93 & 6 & 512 & <0.19 & <1.4\\
 & $^{13}$CO~(1--0) & \nodata & <0.93 & 6 & 510 & <0.19 & <1.4\\
\hline
AM~2055-425 & HCN~(1--0) & 12343 \pm 23 & 0.86 \pm 0.19 & 2 & 211 & 0.18 \pm 0.03 & 1.2 \pm 0.2\\
($\sigma$ = 0.19 mK) & HCO$^{+}$(1--0) & 12324 \pm 12 & 1.60 \pm 0.19 & 3 & 315 & 0.39 \pm 0.03 & 2.7 \pm 0.2\\
 & HNC~(1--0) & \nodata & <0.57 & 3 & 310 & <0.10 & <0.71\\
 & CS~(2--1) & \nodata & <0.57 & 3 & 287 & <0.09 & <0.66\\
 & C$^{18}$O~(1--0) & \nodata & <0.57 & 5 & 427 & <0.11 & <0.76\\
 & $^{13}$CO~(1--0) & 12335 \pm 19 & 1.01 \pm 0.19 & 5 & 425 & 0.27 \pm 0.04 & 1.9 \pm 0.3\\
 & CN~(1--0; 3/2--1/2) & 12299 \pm 37 & 0.62 \pm 0.19 & 4 & 330 & 0.16 \pm 0.03 & 1.1 \pm 0.2\\
 & $^{12}$CO~(1--0) & 12328 \pm 1 & 31.22 \pm 0.19 & 7 & 569 & 6.87 \pm 0.04 & 48.1 \pm 0.3\\
\hline
NGC~7135 & HCN~(1--0) & \nodata & <0.78 & 3 & 317 & <0.14 & <1.0\\
($\sigma$ = 0.26 mK) & HCO$^{+}$(1--0) & \nodata & <0.78 & 3 & 315 & <0.14 & <0.99\\
 & HNC~(1--0) & \nodata & <0.78 & 3 & 310 & <0.14 & <0.98\\
 & CS~(2--1) & \nodata & <0.78 & 3 & 287 & <0.13 & <0.90\\
 & C$^{18}$O~(1--0) & \nodata & <0.78 & 4 & 341 & <0.13 & <0.93\\ 
 & $^{13}$CO~(1--0) & \nodata & <0.78 & 4 & 340 & <0.13 & <0.93\\ 
\hline
NGC~7252 & HCN~(1--0) & 4671 \pm 17 & 1.57 \pm 0.26 & 3 & 317 & 0.33 \pm 0.05 & 2.3 \pm 0.3\\
($\sigma$ = 0.26 mK) & HCO$^{+}$(1--0) & 4616 \pm 30 & 1.02 \pm 0.26 & 3 & 315 & 0.25 \pm 0.05 & 1.8 \pm 0.3\\
 & HNC~(1--0) & \nodata & <0.78 & 3 & 310 & <0.14 & <0.98\\
 & CS~(2--1) & \nodata & <0.78 & 3 & 287 & <0.13 & <0.90\\
 & C$^{18}$O~(1--0) & 4667 \pm 42 & $\sim$0.74\tablenotemark{a} & 4 & 341 & 0.20 \pm 0.04 & 1.4 \pm 0.3\\
 & $^{13}$CO~(1--0) & 4664 \pm 8 & 3.02 \pm 0.26 & 4 & 340 & 0.71 \pm 0.04 & 5.0 \pm 0.3\\
\enddata
\tablecomments{
Column~1: source name and the rms noise level of its spectrum in the antenna temperature unit.
Column~2: molecular line. 
Column~3: the central velocity of the molecular line based on the fitting result for the line identification.
Column~4: the peak antenna temperature. 
Column~5: the number of channels whose values are above 1.5$\sigma$. 
Column~6: the line width. $\Delta\,V_{\rm ch}$ is the velocity width of a channel.
Column~7: the integrated line intensity or its upper limit. 
The errors are estimated from $I_{\rm \sigma} = \sigma \times \Delta\,V_{\rm ch} \sqrt{N_{\rm ch}}$. 
Column~8: the integrated flux density or its upper limit, 
which is estimated from the integrated line intensity 
using a Kelvin-to-Jansky conversion factor of 7 Jy~K($T^{*}_{\rm A}$)$^{-1}$.
}
\tablenotetext{a}{Tentatively-detected line.}
\end{deluxetable*}

\begin{deluxetable*}{lRRRRc}
\tablecaption{Molecular line luminosities\label{tab:t4}}
\tablewidth{0pt}
\tabletypesize{\scriptsize}
\tablehead{
\colhead{Source}&\colhead{$L'_{\rm HCN}$}&\colhead{$L'_{\rm HCO^{+}}$}&\colhead{$L'_{\rm HNC}$}&\colhead{$L'_{\rm CO}$}&\colhead{Reference}\\
&\colhead{(K~km~s$^{-1}$ pc$^{2}$)}&\colhead{(K~km~s$^{-1}$ pc$^{2}$)}&\colhead{(K~km~s$^{-1}$ pc$^{2}$)}&\colhead{(K~km~s$^{-1}$ pc$^{2}$)}
}
\decimalcolnumbers
\startdata
NGC~34 & (1.2\pm0.3)\times10^8 & (1.0\pm0.2)\times10^8 & <4.2\times10^7 & (2.1\pm0.1)\times10^9 & 2\\
NGC~828 & (2.1\pm0.4)\times10^8	& (2.0\pm0.4)\times10^8 & (6.1\pm1.7)\times10^7 & (4.8\pm0.0)\times10^9 & 3\\
UGC~2238 & (1.3\pm0.3)\times10^8 & (1.1\pm0.3)\times10^8 & 	<5.7\times10^7 & <3.8\times10^9& 11\\
NGC~1614 & (1.1\pm0.2)\times10^8 & (1.6\pm0.3)\times10^8 & (3.7\pm1.0)\times10^7 & (2.3\pm0.0)\times10^9 & 4\\
Arp~187 & <2.0\times10^8 & < 1.9\times10^8 & <1.8\times10^8 & (3.8\pm0.2)\times10^9 & 1\\
NGC~2623 & (1.3\pm0.3)\times10^8 & (1.2\pm0.2)\times10^8 & (9.0\pm1.9)\times10^7 & (1.8\pm0.0)\times10^9 & 5\\
NGC~2782 & (1.6\pm0.4)\times10^7 & (2.3\pm0.5)\times10^7 & <5.7\times10^6 & >2.9\times10^8 & 1\\
UGC~5101 & (7.6\pm1.5)\times10^8 & (5.8\pm1.2)\times10^8 & (3.8\pm0.8)\times10^8 & (6.6\pm0.1)\times10^9 & 4\\
AM~0956-282 & <1.0\times10^6 & <1.0\times10^6 & <9.5\times10^5 & (1.1\pm0.1)\times10^7 & 1\\
NGC~3256 & (1.2\pm0.2)\times10^8 & (1.9\pm0.4)\times10^8 & (4.8\pm1.0)\times10^7 & (3.9\pm0.4)\times10^9 & 1\\
NGC~3597 & (1.5\pm0.5)\times10^7 & (2.1\pm0.5)\times10^7 & <9.3\times10^6 & (9.0\pm0.5)\times10^8 & 1\\
NGC~4194 & (1.3\pm0.3)\times10^7 & (2.4\pm0.5)\times10^7 & <6.4\times10^6 & (5.6\pm0.0)\times 10^8 & 4\\
NGC4441 & <6.6\times10^6 & <6.5\times10^6 & <6.2\times10^6 & (1.3\pm0.4)\times10^8 & 6\\
UGC~8058 & (1.5\pm0.3)\times10^9 & (1.4\pm0.3)\times10^9 & (5.0\pm1.1)\times10^8 & (4.7\pm0.5)\times10^9 & 7\\
AM~1300-233	& (9.2\pm2.4)\times10^7	& (1.7\pm0.4)\times10^8 & <4.0\times10^7 & (2.3\pm0.1)\times10^9 & 1\\
Arp~193	& (2.3\pm0.5)\times10^8 & (3.4\pm0.7)\times10^8 & (1.1\pm0.3)\times10^8 & (4.1\pm0.8)\times10^9 & 8\\
UGC~9829 & <7.1\times10^7 & <6.9\times10^7 & <6.6\times10^7 & \nodata & \nodata\\
NGC~6052 & <2.1\times10^7 &	(4.4\pm1.1)\times10^7 & <1.9\times10^7 & (1.1\pm0.0)\times10^9 & 9\\
UGC~10675 & (1.1\pm0.4)\times10^8 & <8.3\times10^7 & <7.9\times10^7 & (9.7\pm0.5)\times10^8 & 10\\
AM~2055-425 & (1.6\pm0.4)\times10^8 & (3.5\pm0.8)\times10^8 & <8.9\times10^7 & (6.5\pm0.3)\times10^9 & 1\\
NGC~7252 & (4.0\pm1.0)\times10^7 & (3.0\pm0.8)\times10^7 & <1.6\times10^7 & (1.1\pm0.1)\times10^9 & 1\\
\enddata
\tablecomments{
Column~1: source name. 
Column~2--4: Molecular line luminosities. The errors are estimated by taking into account 
the rms noise and calibration uncertainties (20\%).
Column~5: The CO luminosity estimated from the integrated intensity measured within 24\arcsec (Reference = 1) 
and literature values of the CO luminosity measured with the IRAM 30 m telescope (Reference = 2--10)
or the the NRAO 12~m telescope (Reference = 11).
Column~6: Reference of the CO data.
}
\tablerefs{
(1) This work; (2) Herrero-Illana et al. (2019); (3) Bertram et al. (2006); 
(4) Costagliola et al. (2011); (5) Garc{\'\i}a-Burillo et al. (2012); (6) J{\"u}tte et al. (2010); 
(7) Aladro et al. (2015); (8) \citet{1997ApJ...478..144S}; (9) \citet{2005ApJ...624..714G}; 
(10) \citet{1999AJ....118..145Z}; (11) \citet{1991ApJ...370..158S}; 
%(1) This work; (2) \citet{2019A&A...628A..71H}; (3) \citet{2006A&A...448...29B}; 
%(4) \citet{1991ApJ...370..158S}; (5) \citet{2011A&A...528A..30C}; (6) \citet{2012A&A...539A...8G}; 
%(7) \citet{2014ApJS..214....1U}; (8) \citet{2010A&A...509A..19J}; (9) \citet{2015A&A...579A.101A}; 
%(10) \citet{1997ApJ...478..144S}; (11) \citet{2005ApJ...624..714G}; (12) \citet{1999AJ....118..145Z}
}
\end{deluxetable*}

\begin{deluxetable*}{lRRc}
\tablecaption{Mean luminosity ratios of the dense gas tracers\label{tab:t5}}
\tablewidth{0pt}
\tabletypesize{\scriptsize}
\tablehead{
\colhead{Samples}&
\colhead{$L'_{\rm HCN}/L'_{\rm HCO^{+}}$}&
\colhead{$L'_{\rm HNC}/L'_{\rm HCN}$}&
\colhead{Reference}
}
\startdata
Merger remnants & 0.88\pm0.08 & 0.43\pm0.05 & 1\\
Late-stage mergers & 1.02\pm0.15 & 0.52\pm0.03 & 2\\
Early/Mid-stage mergers & 0.92\pm0.12 & 0.60\pm0.09 & 2\\
Non-merging LIRGs & 1.38\pm0.11 & 0.56\pm0.07 & 2\\
\enddata
\tablerefs{
(1) This work; (2) \citet{2015ApJ...814...39P}
}
\end{deluxetable*}

\begin{deluxetable*}{lRRccRRc}
\tablecaption{Dense gas fraction, SFR$_{\rm dense}$, and SFE\label{tab:t6}}
\tablewidth{0pt}
\tabletypesize{\scriptsize}
\tablehead{
\colhead{Samples}&
\twocolhead{$L'_{\rm HCN}/L'_{\rm CO}$}&
\twocolhead{$L_{\rm IR}/L'_{\rm HCN}$}&
\twocolhead{$L_{\rm IR}/L'_{\rm CO}$}&
\colhead{Reference}\\
&\colhead{mean}&\colhead{median}&\colhead{mean}&\colhead{median}&\colhead{mean}&\colhead{median}&
}
\startdata
Merger remnants & 0.070$\pm$0.020 & 0.045 & $(2.9\pm0.5) \times 10^{3}$ & $2.4 \times 10^{3}$ & 150$\pm$30 & 123 & 1\\
Late-stage mergers & 0.093$\pm$0.021 & 0.063 & $(2.0\pm0.3) \times 10^{3}$ & $2.1 \times 10^{3}$ & 170$\pm$40 & 117 & 2,3\\
Early/Mid-stage mergers & 0.093$\pm$0.027 & 0.050 & $(1.6\pm0.5) \times 10^{3}$ & $1.3 \times 10^{3}$ & 87$\pm$17 & 65 & 2,3\\
Non-merging LIRGs & 0.098$\pm$0.012 & 0.078 & $(8.5\pm0.6) \times 10^{2}$ & $8.5 \times 10^{2}$ & 89$\pm$12 & 72 & 2,3\\
Late-type galaxies & 0.061$\pm$0.008 & 0.049 & $(7.5\pm1.0) \times 10^{2}$ & $5.1 \times 10^{2}$ & 39$\pm$6 & 28 & 4\\
\enddata
\tablerefs{
(1) This work; (2) \citet{2015ApJ...814...39P}; (3) Herrero-Illana et al. (2019); (4) \citet{2004AJ....128.2098R}
}
\end{deluxetable*}

\begin{deluxetable*}{lRRR}
\tablecaption{Kolmogorov-Smirnov test $P$-value\label{tab:t7}}
\tablewidth{0pt}
\tabletypesize{\scriptsize}
\tablehead{
\colhead{Samples}&
\colhead{$L'_{\rm HCN}/L'_{\rm CO}$}&
\colhead{$L_{\rm IR}/L'_{\rm HCN}$}&
\colhead{$L_{\rm IR}/L'_{\rm CO}$}
}
\startdata
Merger remnants -- Late-stage mergers & 0.38 & 0.53 & 0.95\\
Merger remnants -- Early/Mid-stage mergers & 0.81 & 0.03 & 0.04\\
Merger remnants -- Non-merging LIRGs & $\ll$0.05 & $\ll$0.05 & 0.03\\
Merger remnants -- Late-type galaxies & 0.44 & $\ll$0.05 & $\ll$0.05\\
\enddata
\end{deluxetable*}

\begin{figure*}[htbp]
	\begin{center}
		\includegraphics[width=0.9\textwidth, trim=0cm 0cm 0cm 0cm, clip]{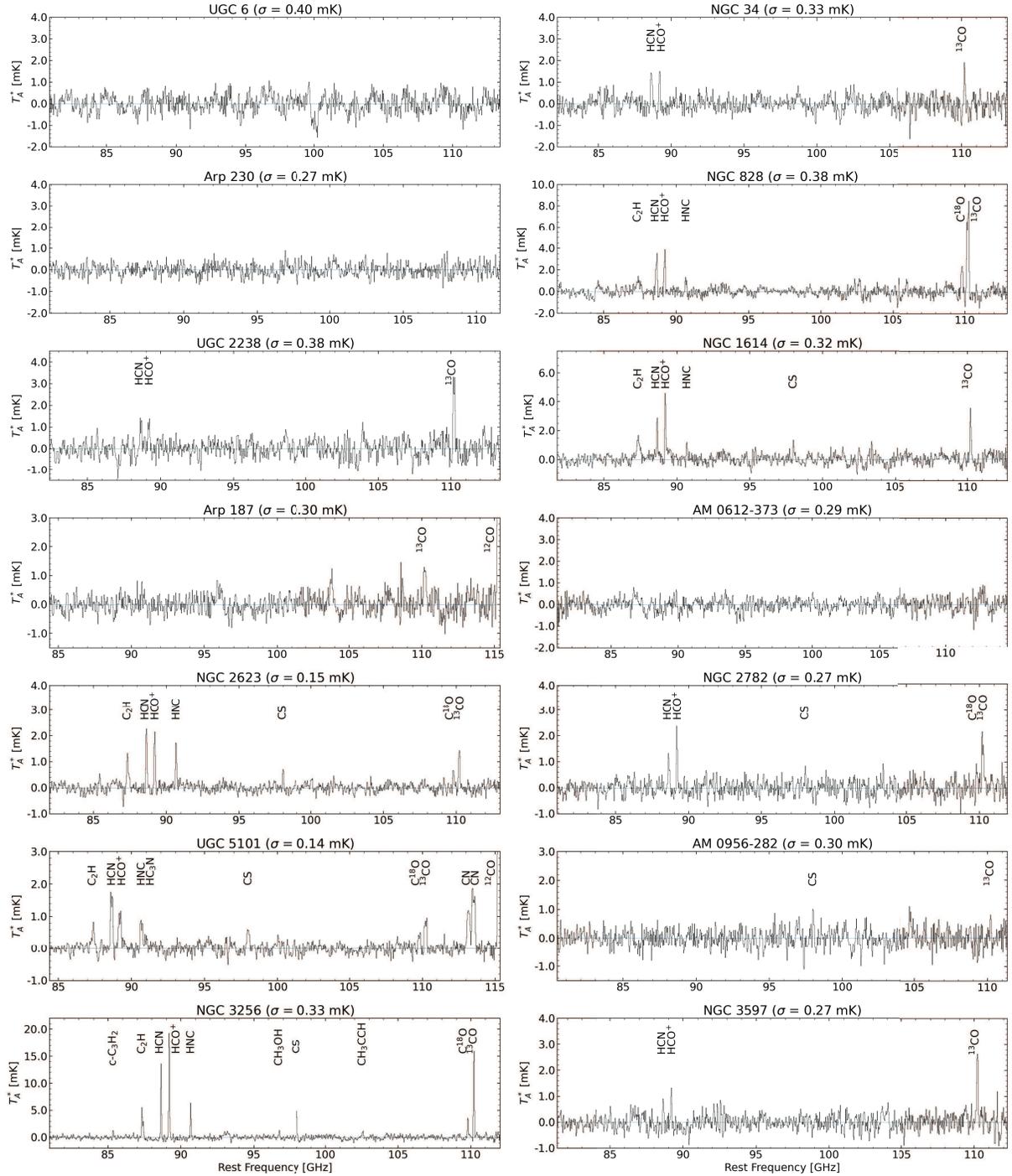}
	\end{center}
	\caption{
	Spectra of the merger remnants.
	The intensity scale is in the antenna temperature ($T_{\rm A}^{*}$). 
	The frequency resolution is $\sim$31~MHz.
	\label{fig:f1}
	}
\end{figure*}

\addtocounter{figure}{-1}
\begin{figure*}[htbp]
	\begin{center}
		\includegraphics[width=0.9\textwidth, trim=0cm 0cm 0cm 0cm, clip]{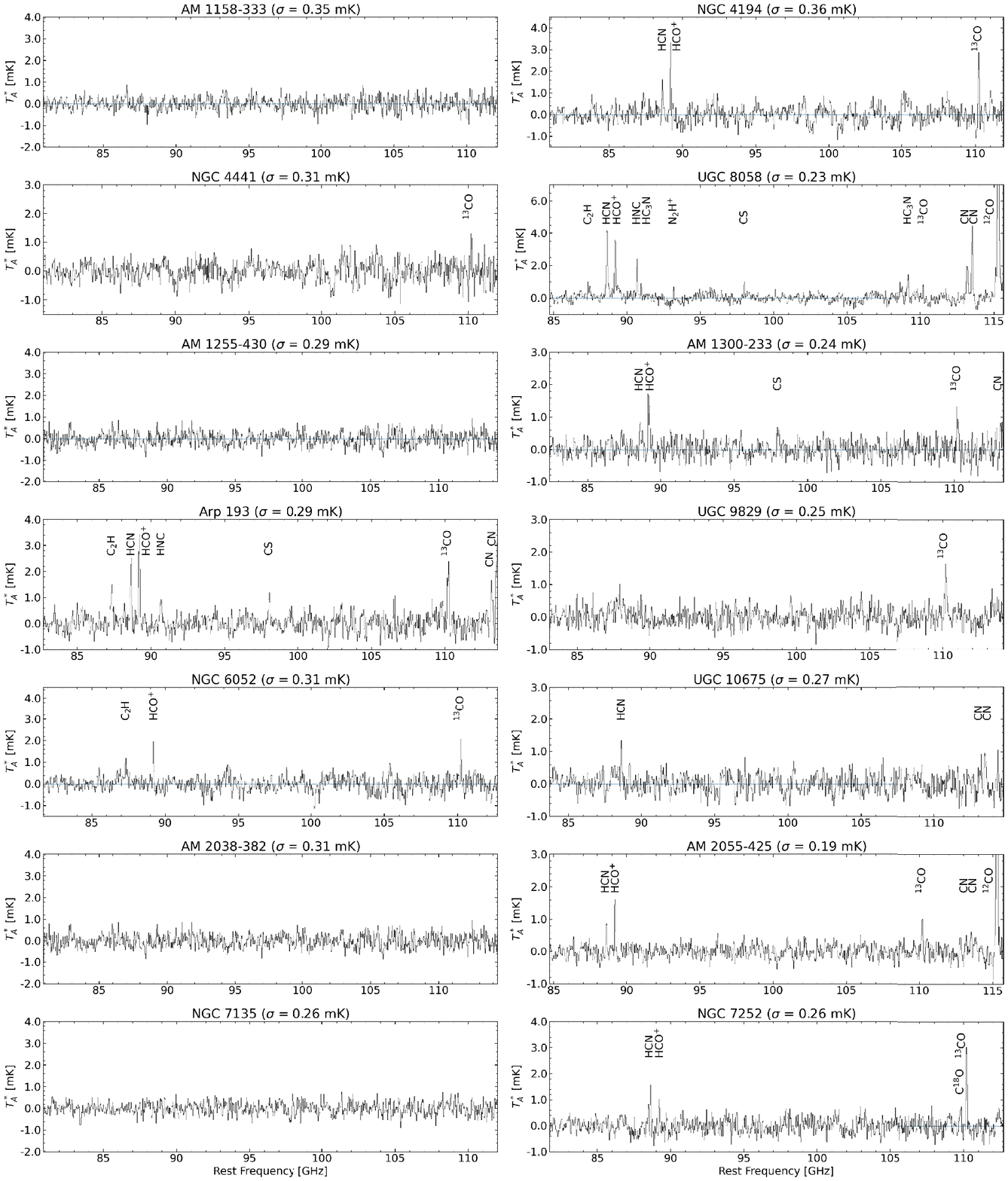}
	\end{center}
	 \caption{
	 Continued.
	 }
\end{figure*}

\begin{figure*}[htbp]
	\begin{center}
		\includegraphics[width=1.0\textwidth]{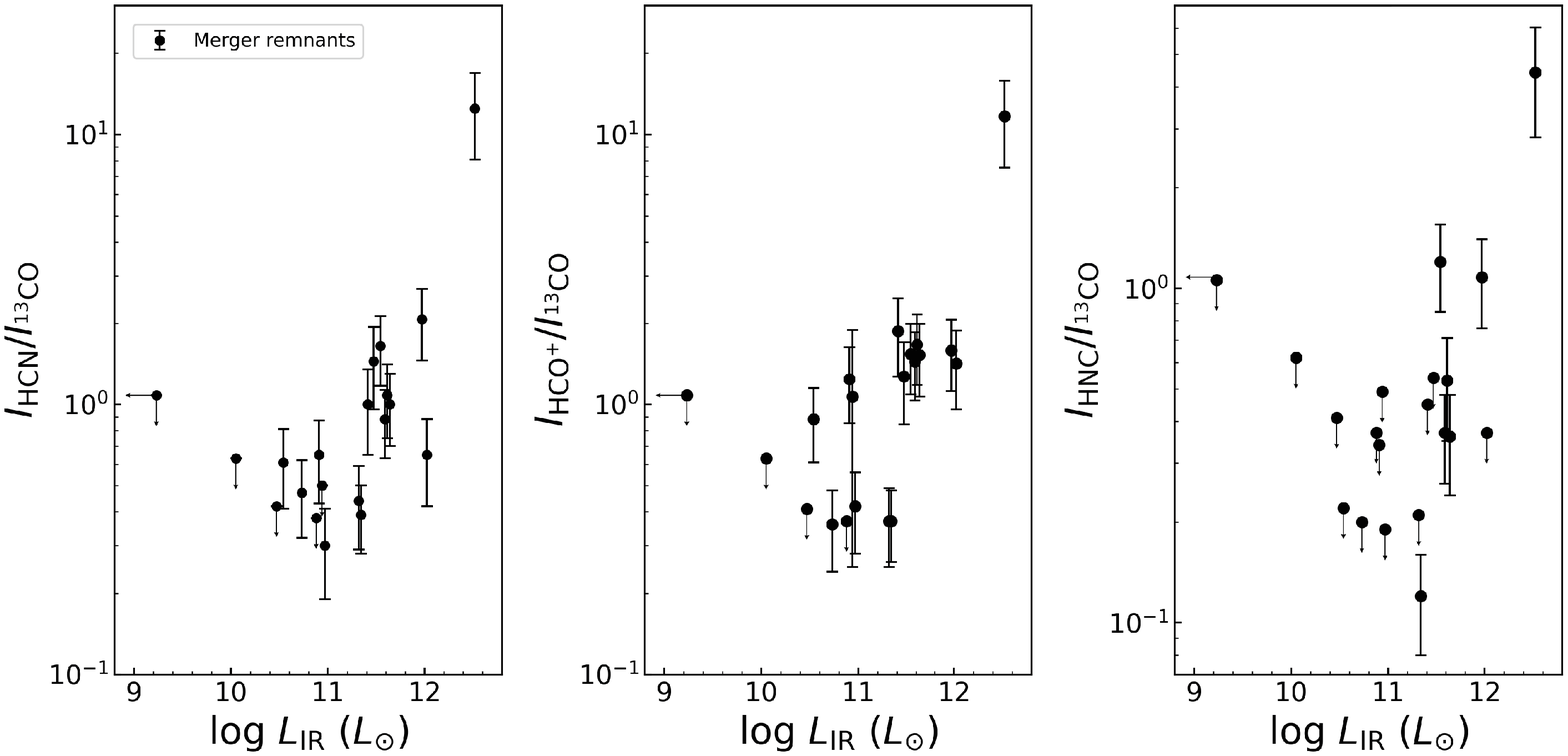}
	\end{center}
	\caption{
	Plot of the HCN~(1--0)/$^{13}$CO~(1--0), HCO$^{+}$(1--0)/$^{13}$CO~(1--0), 
	and HNC~(1--0)/$^{13}$CO~(1--0) intensity ratios as a function of the IR luminosity.
	\label{fig:f2}
	}
\end{figure*}

\begin{figure*}[htbp]
	\begin{center}
		\includegraphics[width=1.0\textwidth]{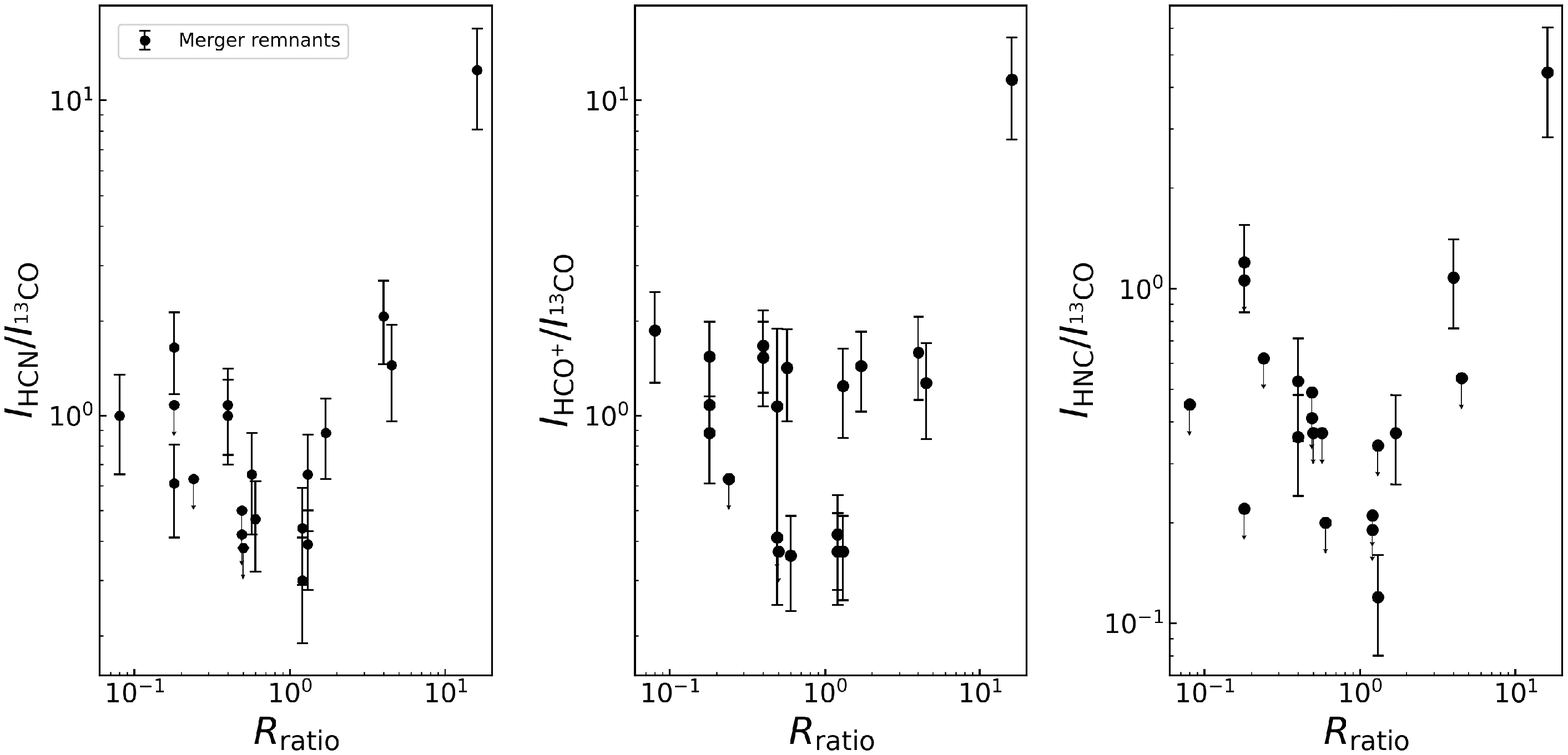}
	\end{center}
	\caption{
	The same as Figure~\ref{fig:f2}, but the x-axis is the size of the molecular gas disk 
	relative to the stellar component ($R_{\rm ratio}$; see the text for details). 
	\label{fig:f3}
	}
\end{figure*}

\begin{figure*}[htbp]
	\begin{center}
		\includegraphics[width=1.0\textwidth]{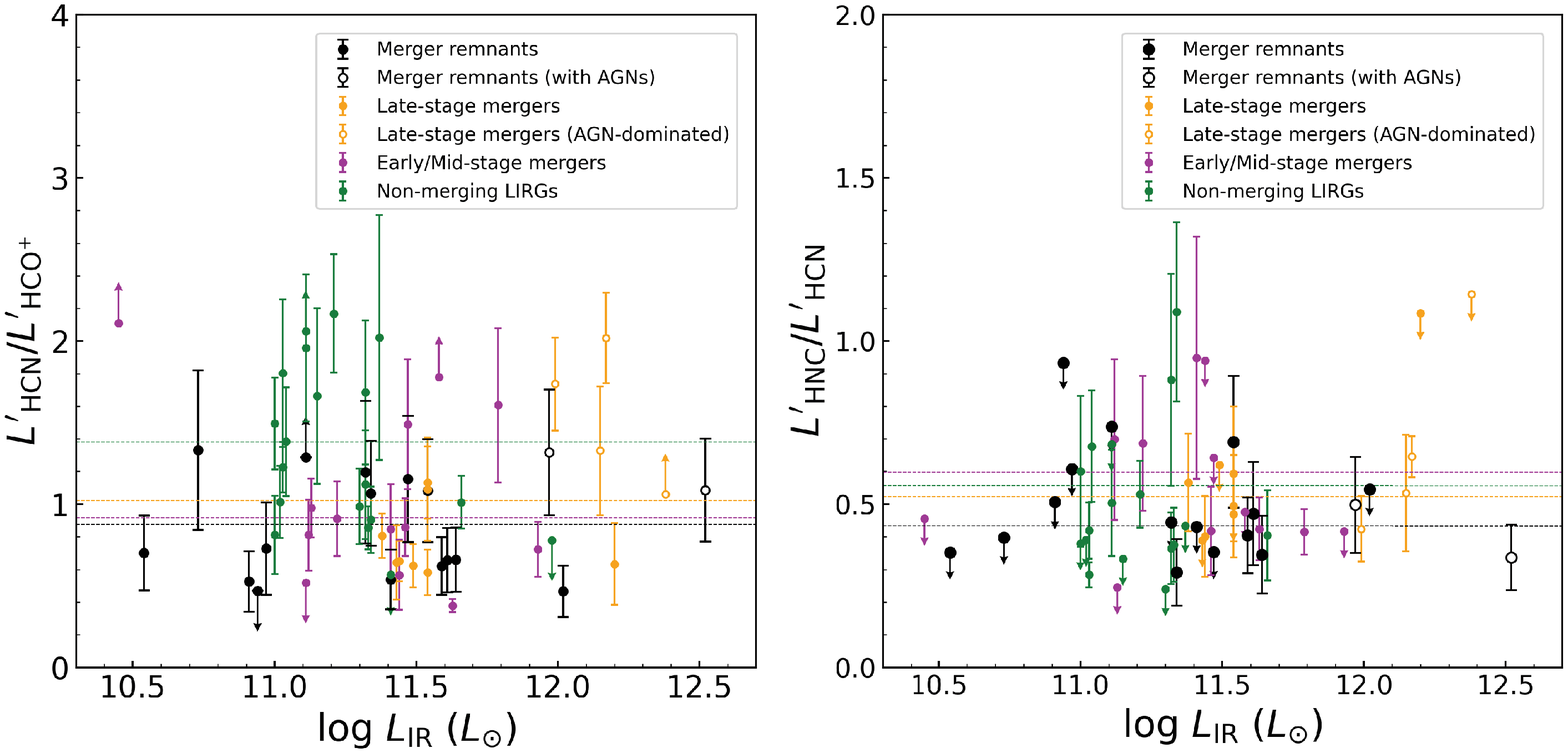}
	\end{center}
	\caption{
	(left) Plot of the HCN~(1--0)/HCO$^{+}$(1--0) luminosity ratios as a function of the IR luminosity.
	The black circles show the merger remnants.
	The orange, magenta, and green circles show the late-stage mergers, early/mid-stage mergers, 
    and non-merging LIRGs, respectively, in the subsample of the GOALS \citep{2015ApJ...814...39P}.
	The open black circles and open orange circles show sources which host AGNs or AGN-dominated sources.
	Each dotted line presents the average HCN/HCO$^{+}$ luminosity ratio in each sample.
	(right) Plot of the HNC~(1--0)/HCN~(1--0) luminosity ratios as a function of the IR luminosity.  
	The symbols and dotted lines are the same as those in the left figure.
	\label{fig:f4}
	}
\end{figure*}

\begin{figure*}[htbp]
	\begin{center}
		\includegraphics[width=1.0\textwidth]{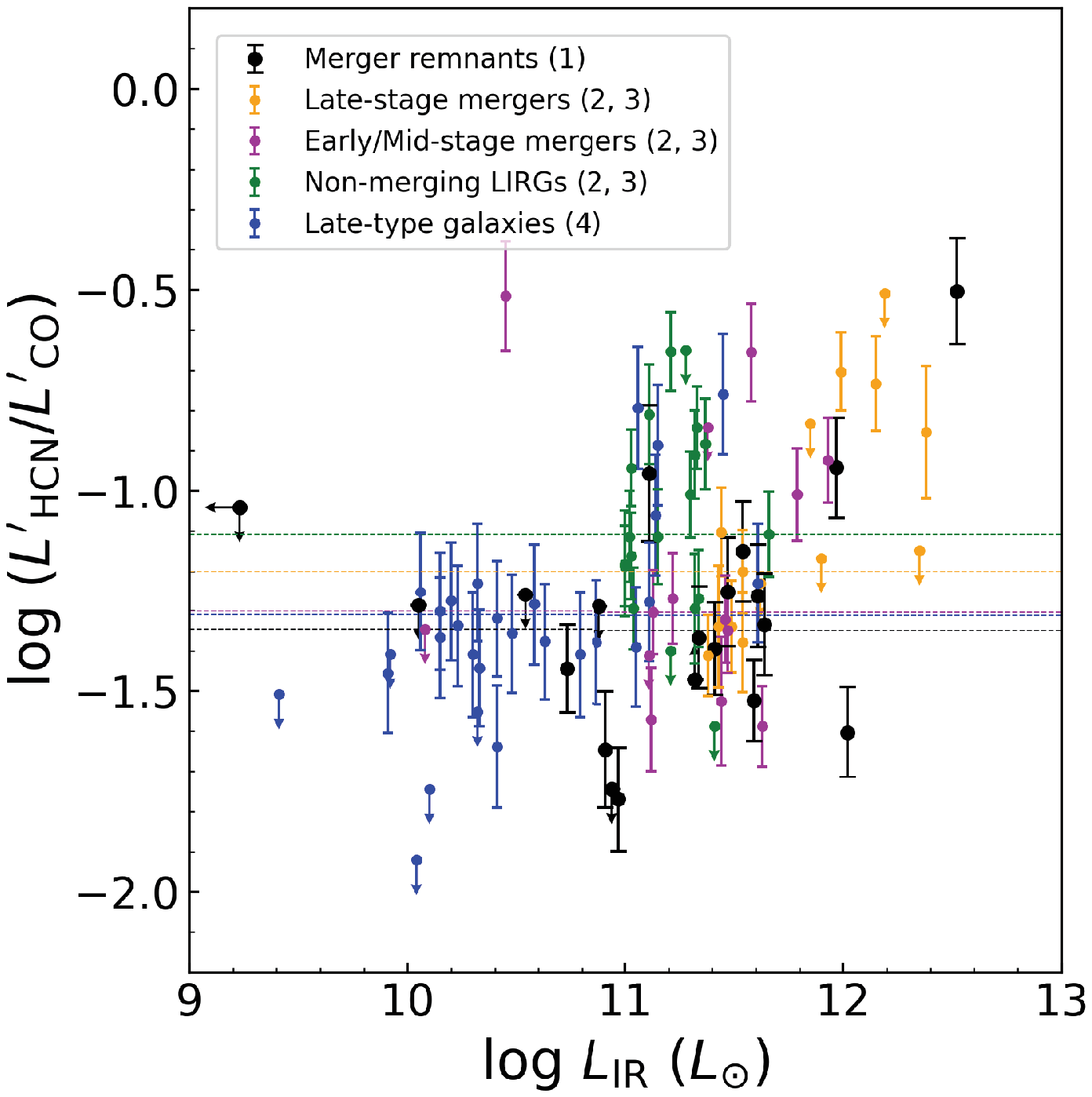}
	\end{center}
	\caption{
	Plot of the HCN~(1--0)/CO~(1--0) luminosity ratios as a function of the IR luminosity.
	The black circles show the merger remnants.
	The orange, magenta, and green circles show late-stage mergers, early/mid-stage mergers, 
	and non-merging LIRGs, respectively, in the GOALS sample.
	The blue circles show late-type galaxies.
	Each dotted line shows the median $L'_{\rm HCN}/L'_{\rm CO}$ of each sample.
	The numbers in the legend are references: (1) this work, (2) \citet{2015ApJ...814...39P}, 
	(3) \citet{2019A&A...628A..71H}, (4) \citet{2004ApJ...606..271G}
	\label{fig:f5}
	}
\end{figure*}

\begin{figure*}[htbp]
	\begin{center}
		\includegraphics[width=1.0\textwidth]{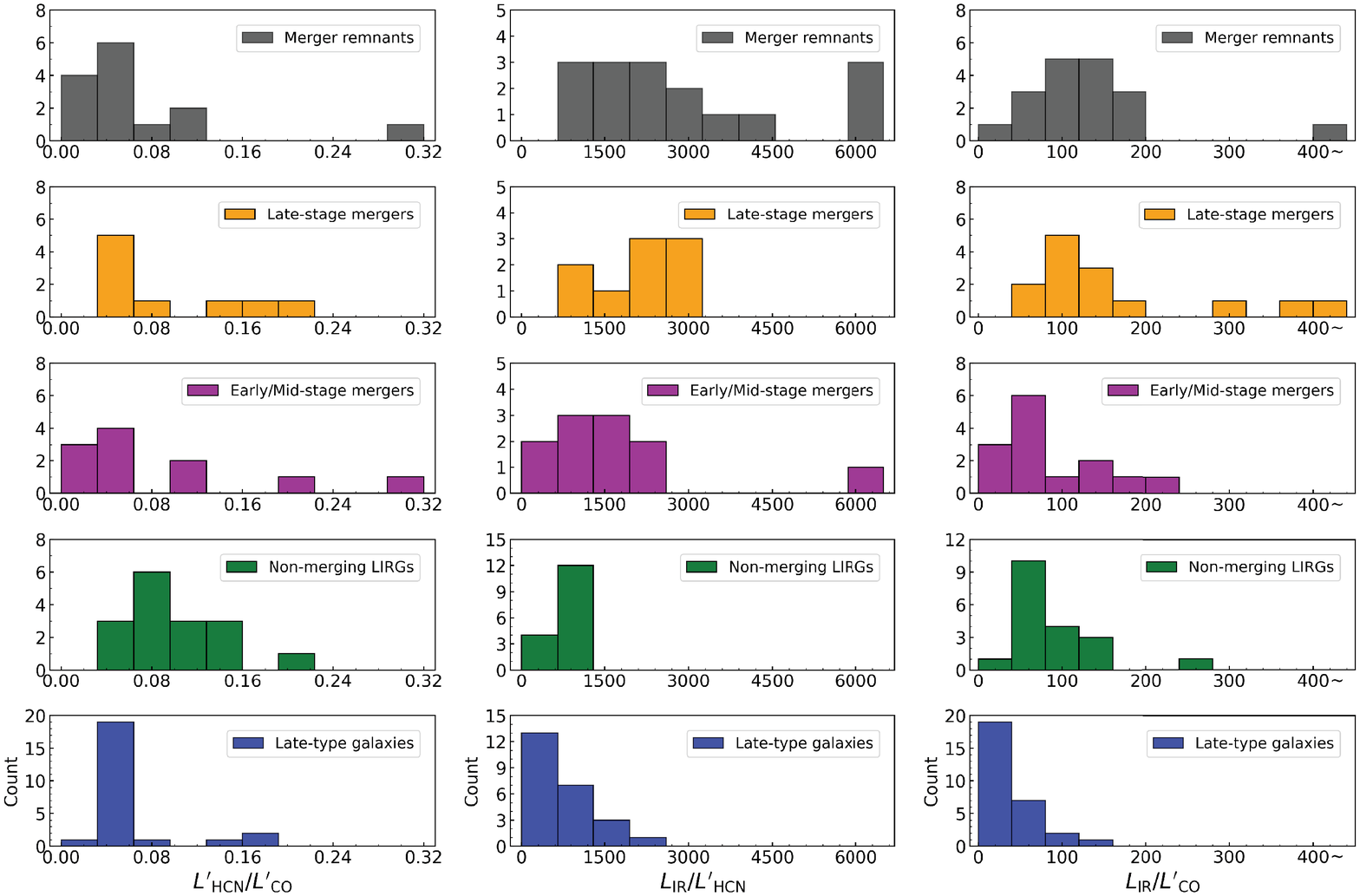}
	\end{center}
	\caption{
	Histograms of the HCN~(1--0)/CO~(1--0) luminosity ratio (left), 
	the IR-to-CO~(1--0) luminosity ratio (middle), and the IR-to-HCN~(1--0) luminosity ratio (right).
	The different colors show the different samples.
	\label{fig:f6}
	}
\end{figure*}

\begin{figure*}[htbp]
	\begin{center}
		\includegraphics[width=1.0\textwidth]{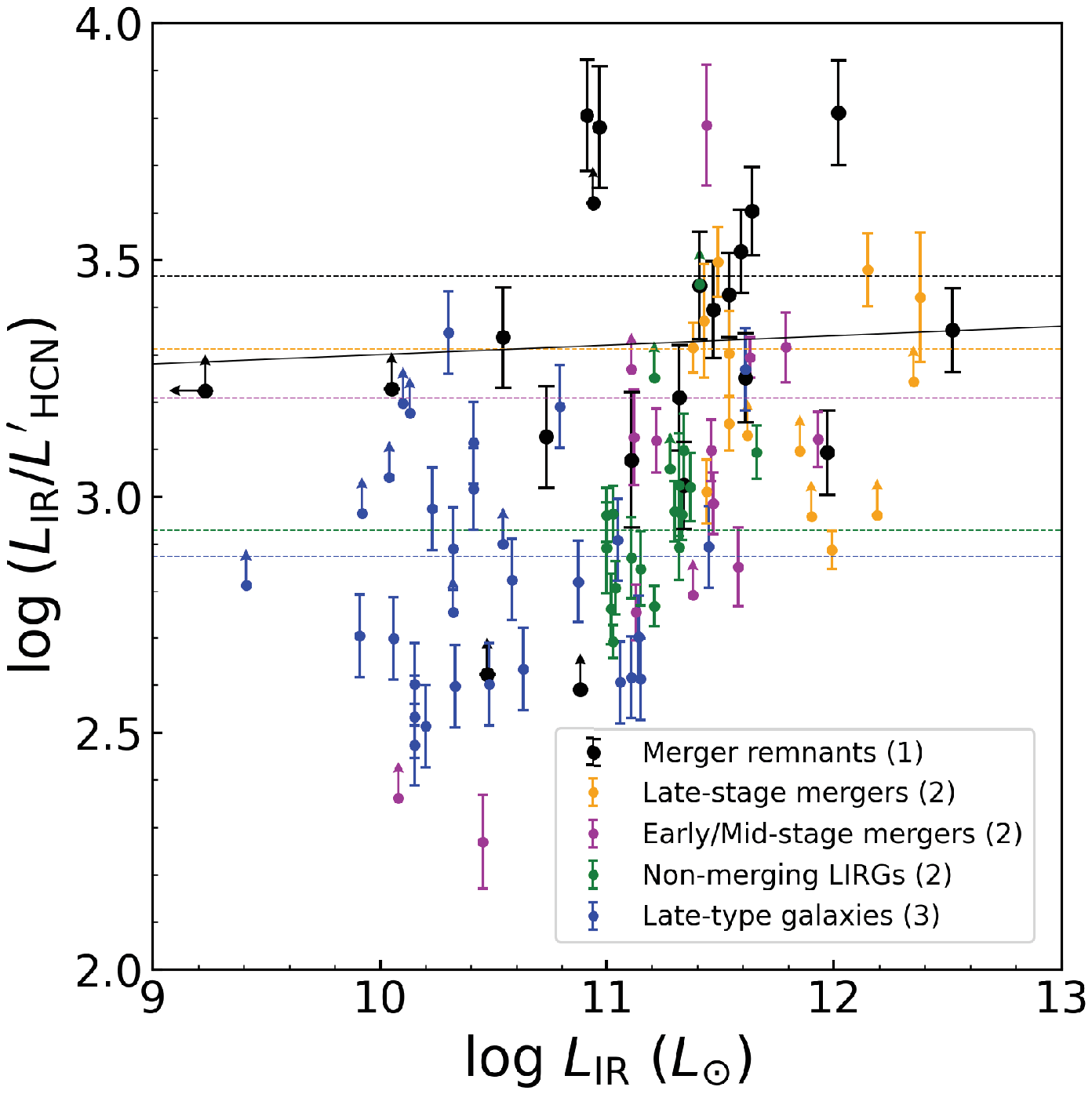}
	\end{center}
	\caption{
	Plot of the IR-to-HCN~(1--0) luminosity ratios as a function of the IR luminosity.
	The symbols are the same as Figure~\ref{fig:f5}.
	Each dotted line shows the mean $L_{\rm IR}/L'_{\rm HCN}$ of each sample.
	The solid black line is the best-fit line for the merger remnant sample.
	The numbers in the legend are references: (1) this work, 
	(2) \citet{2019A&A...628A..71H}, (3) \citet{2004ApJ...606..271G}
	\label{fig:f7}
	}
\end{figure*}

\begin{figure*}[htbp]
	\begin{center}
		\includegraphics[width=1.0\textwidth]{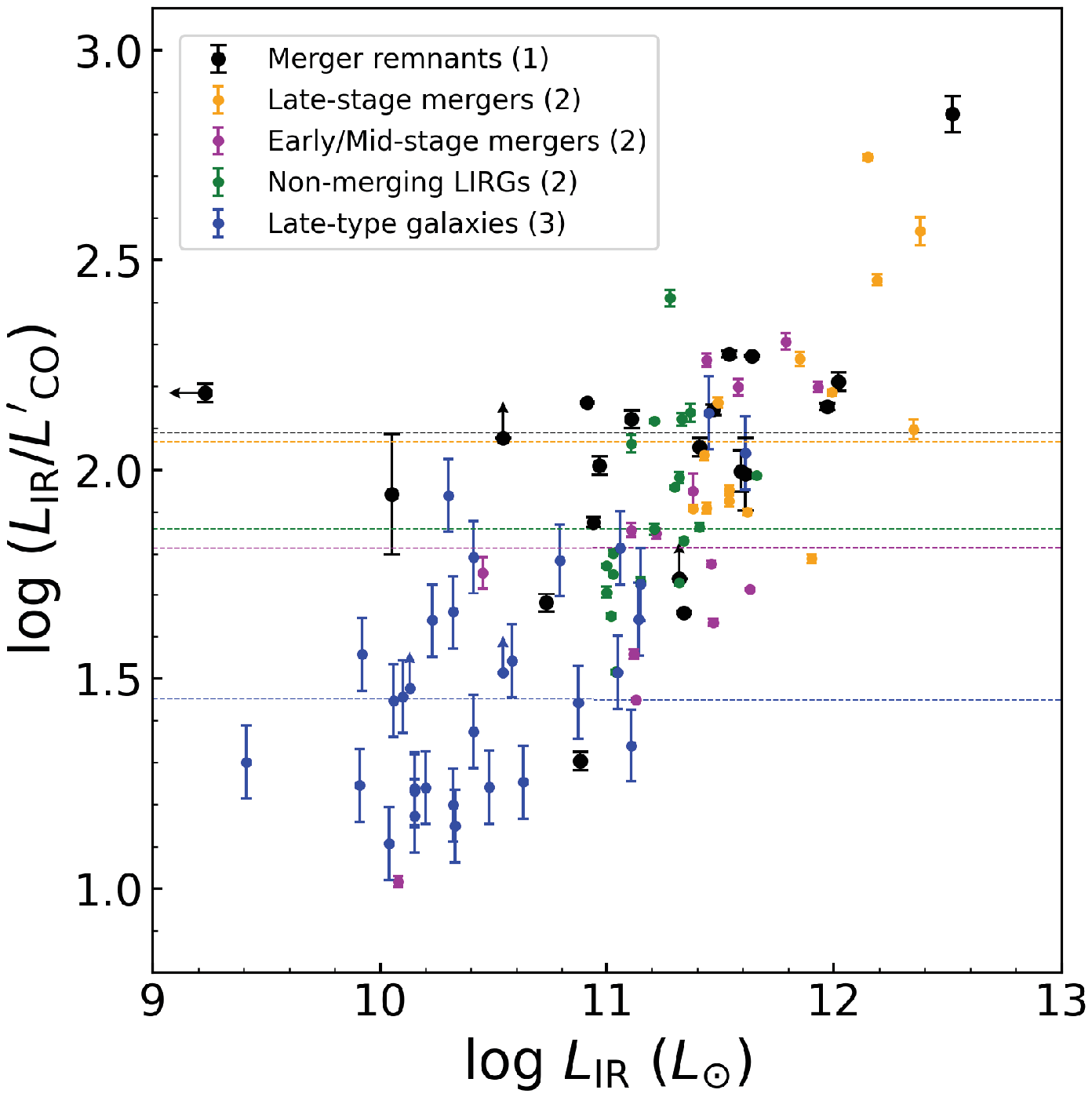}
	\end{center}
	\caption{
	Plot of the IR-to-CO~(1--0) luminosity ratios as a function of the IR luminosity.
	The symbols are the same as Figure~\ref{fig:f5}.
	Each dotted line shows the median $L_{\rm IR}/L'_{\rm CO}$ of each sample.
	The numbers in the legend are references: (1) this work, 
	(2) \citet{2015ApJ...814...39P}, (3) \citet{2004ApJ...606..271G}
	\label{fig:f8}
	}
\end{figure*}

\begin{figure*}[htbp]
	\begin{center}
		\includegraphics[width=1.0\textwidth]{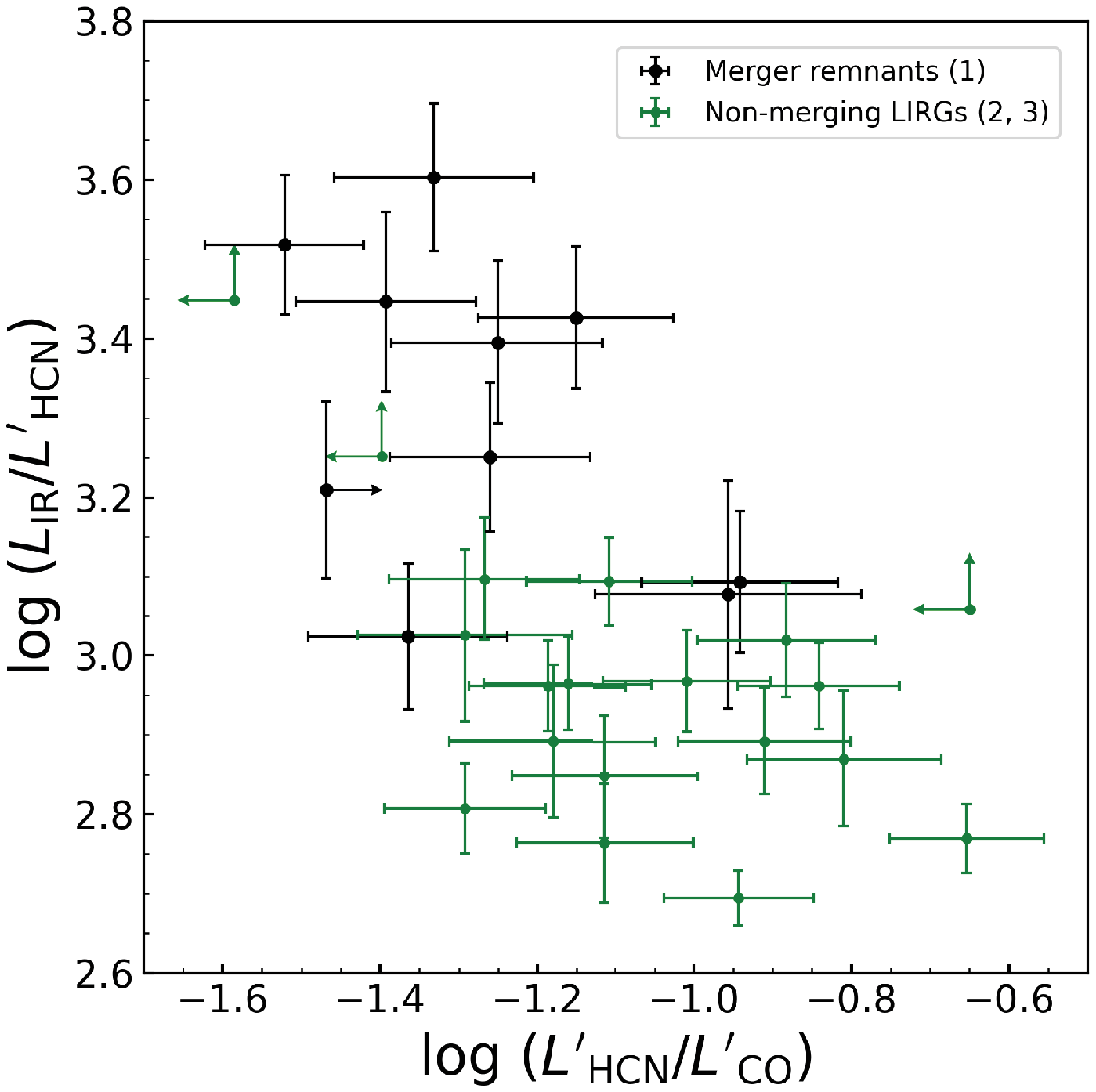}
	\end{center}
	\caption{
	Plot of the IR-to-HCN~(1--0) luminosity ratios 
	as a function of the HCN~(1--0)/CO~(1--0) luminosity ratio.
	The black circles are the merger remnants which are classified as LIRGs, 
	and the green circles are the non-merging LIRGs.
	The numbers in the legend are references: (1) this work, 
	(2) \citet{2015ApJ...814...39P}, (3) \citet{2019A&A...628A..71H}
	\label{fig:f9}
	}
\end{figure*}

\clearpage
\bibliography{ms}{}
\bibliographystyle{aasjournal}

\listofchanges

\end{document}